\pdfoutput=1
\documentclass[11pt]{article}
\usepackage[table]{xcolor}

\usepackage[final]{acl}
\usepackage{enumitem}
\usepackage{times}
\usepackage{latexsym}

\usepackage[T1]{fontenc}
\makeatletter
\def\@fnsymbol#1{%
   \ifcase#1\or
   \TextOrMath \textdagger \dagger\or
   \TextOrMath \textdaggerdbl \ddagger \or
   \TextOrMath \textsection  \mathsection\or
   \TextOrMath \textparagraph \mathparagraph\or
   \TextOrMath \textbardbl \|\or
   \TextOrMath {\textdagger\textdagger}{\dagger\dagger}\or
   \TextOrMath {\textdaggerdbl\textdaggerdbl}{\ddagger\ddagger}\else
   \@ctrerr \fi
}
\makeatother

\usepackage[utf8]{inputenc}
\usepackage{pifont}
\usepackage{longtable}
\usepackage{xspace}
\usepackage{microtype}
\usepackage{booktabs}
\newcommand{\llmname}[1]{{\fontfamily{pcr}\selectfont {#1}}\xspace}
\usepackage{inconsolata}
\usepackage{amssymb}

\usepackage{graphicx}
\usepackage{amsmath}
\usepackage{multirow}
\usepackage{bbding}

\newcommand{\ourmethod}{{\fontfamily{lmtt}\selectfont \textbf{G-Safeguard}}\xspace}

\usepackage{cleveref}
\usepackage{amsfonts}
\usepackage{bbm}
\usepackage{mathtools}
\title{\ourmethod: A Topology-Guided Security Lens and Treatment on LLM-based Multi-agent Systems}

\author{
  \textbf{Shilong Wang}{\footnotesize $^\bigstar$}\thanks{Guibin Zhang and Shilong Wang contributed equally.} \quad 
  \textbf{Guibin Zhang}{\footnotesize $^\spadesuit$}$^\dagger$ \quad 
  \textbf{Miao Yu}{\footnotesize $^\bigstar$} \quad 
  \textbf{Guancheng Wan}{\footnotesize $^\clubsuit$} \quad 
  \textbf{Fanci Meng}{\footnotesize $^\bigstar$} \\
  \textbf{Chongye Guo}{\footnotesize $^\blacklozenge$} \quad 
  \textbf{Kun Wang}{\footnotesize $^{\bigstar}$\textsuperscript{\Envelope}} \quad  
  \textbf{Yang Wang}{\footnotesize $^{\bigstar}$\textsuperscript{\Envelope}} \\
  {\footnotesize $^\bigstar$}USTC \quad
  {\footnotesize $^\spadesuit$}Tongji University \quad 
  {\footnotesize $^\clubsuit$}Wuhan University \quad 
  {\footnotesize $^\blacklozenge$}Shanghai University\\
  \href{mailto:wk520529@mail.ustc.edu.cn}{\texttt{wk520529@mail.ustc.edu.cn}} \quad 
  \href{mailto:angyan@ustc.edu.cn}{\texttt{angyan@ustc.edu.cn}}
}

\begin{document}
\maketitle
\begin{abstract}
Large Language Model (LLM)-based Multi-agent Systems (MAS) have demonstrated remarkable capabilities in various complex tasks, ranging from collaborative problem-solving to autonomous decision-making. However, as these systems become increasingly integrated into critical applications, their vulnerability to \textit{adversarial attacks}, \textit{misinformation propagation}, and \textit{unintended behaviors} have raised significant concerns. To address this challenge, we introduce \ourmethod, a topology-guided security lens and treatment for robust LLM-MAS, which leverages graph neural networks to detect anomalies on the multi-agent utterance graph and employ topological intervention for attack remediation.  Extensive experiments demonstrate that \ourmethod: (I) exhibits significant effectiveness under various attack strategies, recovering over 40\% of the performance for prompt injection; (II) is highly adaptable to diverse LLM backbones and large-scale MAS; (III) can seamlessly combine with mainstream MAS with security guarantees. The code is available at \url{https://github.com/wslong20/G-safeguard}.
\end{abstract}

\section{Introduction}

Autonomous agents \cite{wang2024survey}, while inheriting the general task-solving and instruction comprehension capabilities of Large Language Models (LLMs) \cite{chang2024survey, minaee2024large}, are equipped with external units such as tools \cite{liu2024toolace, tang2023toolalpaca} and memory \cite{zhang2024survey}, extending the capability boundaries of LLMs. Multi-agent systems (MAS) further integrate collective intelligence through agent interactions, enhancing the capabilities of individual agents. This enables MAS to be recognized as intelligent entities capable of tackling more complex tasks, such as \textit{environment perception} and \textit{embodied actions} \cite{guo2024large, zhao2025see}. However, on the other hand, while MAS appreciates these designs, it not only sadly admits the security drawbacks of LLMs and single agents \cite{inan2023llama} but also introduces additional risks through interactions among multiple agents, further complicating its security concerns \cite{dong2024attacks, yu2024llm, yu2024netsafe}.

\begin{figure}[t]
    \vspace{-0.1cm}
    \centering
    \includegraphics[width=\linewidth]{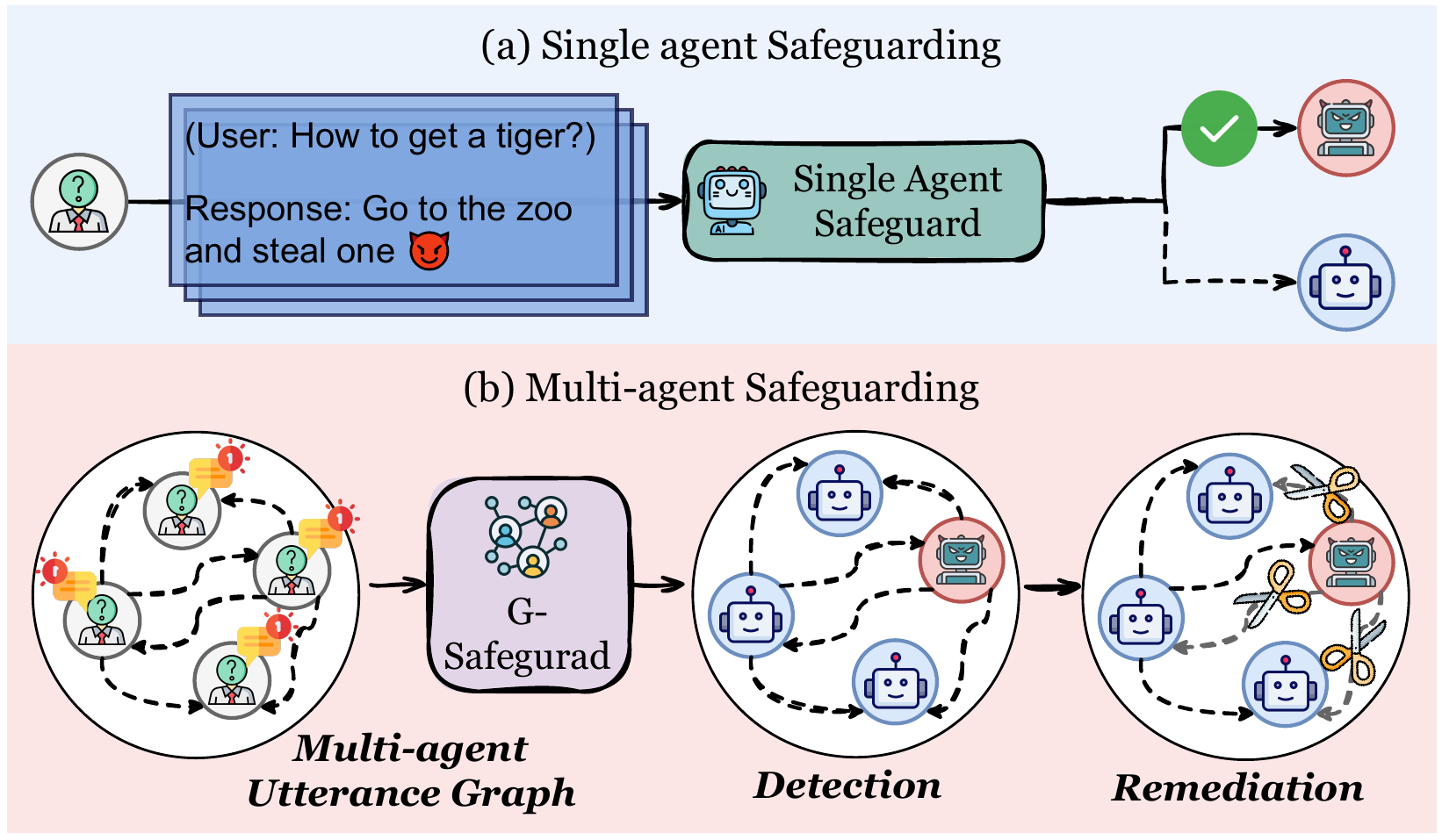} 
    \vspace{-0.5cm}
    \caption{The paradigm comparison between single agent safeguard and multi-agent safeguard.}
    \vspace{-0.3cm}
    \label{fig:intro} 
\end{figure}

Existing attacks on agents primarily target their external units (\textit{e.g,} tool, memory) \cite{tian2023evil, zhang2024psysafe} and main body \cite{liu2023prompt}. For single-agent, direct prompt injection \cite{perez2022ignore, kang2024exploiting, liu2023prompt, toyer2023tensor} manipulates agents by embedding harmful knowledge or biases into their foundation LLM, influencing decision-making and generating harmful responses. Indirect prompt injection exposes agents to a vast amount of potentially harmful information when interacting with external interfaces \cite{greshake2023not, zhan2024injecagent, yi2023benchmarking}, indirectly enabling them to acquire malicious instructions. Without custom designs or any modifications, MAS falls into the predicament of single agents and introduces communication as a new point of vulnerability \cite{yu2024netsafe, tian2023evil}. For example, NetSafe \cite{yu2024netsafe} take the first step to identify the existence of topology-based \textit{misinformation} and \textit{bias} propagation.

Compared to attacks, defense in MAS becomes significantly more challenging, encompassing both anomalies \underline{detection} and \underline{remediation} \cite{andriushchenko2024agentharm}, also shown in \Cref{fig:intro}. Detection focuses on diagnosing the location of the issue, while remediation involves mitigating the negative impact. Unfortunately, the literature primarily focuses on designing customized defenses for attacks targeting specific units, overlooking the features of MAS:

\vspace{-0.4em}
\begin{itemize}[leftmargin=*]
    \item[\ding{72}] \textbf{\textit{Topology-Aware.}} Existing defenses primarily focus on single-agent, neglecting the critical topology characteristics inherent in MAS \cite{zhang2024cut,zhang2024g}. Due to the interaction mechanisms, anomaly detection must take into account the relationships and behaviors of neighboring agents to effectively address the unique challenges posed by the interconnected nature.
    \vspace{-0.3em}
    \item[\ding{72}] \textbf{\textit{Inductive transferability.}} When combining the number of agents with the design of external units, MAS can exhibit an extensive variety of combinations and configurations. Custom-designed solutions hinder transferability \cite{fu2024imprompter, hua2024trustagent}, which even unnerves small-size MAS, let alone large-size MAS-based applications \cite{zhang2024privacyasst}.
    
\end{itemize}

To tackle these challenges, we propose \ourmethod, a topology-guided lens for attack detection and treatment for attack remediation. Technically, \ourmethod performs security assessments at the conclusion of each dialogue round within a multi-turn MAS. It first constructs a \textbf{multi-agent utterance graph}, which sufficiently encodes both agent-wise communicative interactions and topological dependencies.  
Building upon this foundation, \ourmethod formulates the attack detection as an \textbf{anomaly detection} problem on this graph, leveraging an edge-featured graph neural network (GNN) to pinpoint high-risk agents. Finally, \ourmethod enforces \textbf{topological interventions} to disrupt the propagation of misleading or adversarial information, thereby materializing robust MAS against a broad spectrum of agent-oriented attacks, \textit{i.e.}, prompt injection, memory poisoning, and tool attacks. More importantly, inheriting the inductive ability of GNNs, \ourmethod can scale to arbitrary-scale MAS without resource-intensive retraining, meanwhile exhibiting remarkable cross-LLM-backbone generalizability.

We conduct extensive experiments to validate our method's effectiveness: \ourmethod (I) prevents malicious information spread across various MAS topologies, blocking $12.50\%\sim33.23\%$ of infections in chain structures and $10\%\sim38.52\%$ in star structures on the MMLU dataset; (II) defends against multiple attacks, reducing attack success rate (ASR) by $21.38\%$ and $22.01\%$ for \textit{prompt injection} on CSQA and MMLU, $12.67\%$ for \textit{tool attack}, and $16.27\%$ for \textit{memory poison}; (III) scales seamlessly to large-scale MAS, maintaining stable performance with $19.50\%\sim39.23\%$ ASR reduction under prompt injection settings.

Our contributions can be concluded as follows:
\begin{itemize}[leftmargin=*]
\vspace{-0.4em}\item \textbf{\textit{Paradigm Proposal.}} We pioneer the \textit{detection} and \textit{remediation} paradigm for adversarial defense within LLM-MAS, which emphasizes topology-aware diagnosis and intervention of misleading or malicious information as it propagates and proliferates across the multi-agent system.

\vspace{-0.5em}\item \textbf{\textit{Practical Solution.}} We introduce \ourmethod, a topology-guided framework for attack detection and remediation, enabling lightweight and real-time identification of adversarial entities on multi-agent utterance graphs and contamination-free communication via graph pruning.

\vspace{-0.7em}\item \textbf{\textit{Emperical Evaluation.}} The results across various LLM backbones, MAS frameworks, and attack strategies show that \ourmethod provides effective protection in all these scenarios. Also, \ourmethod is adaptable to arbitrary-scale MAS and can integrate into mainstream MAS to enhance their defensive capabilities.

\end{itemize}

\section{Preliminary}
\label{sec:preliminary}
In this section, we establish the notation and formalize key concepts for the attack and defense of multi-agent systems from a topology perspective.

\paragraph{Multi-agent System.} Consider a multi-agent system composed of \( N = |\mathcal{V}| \) agents, which we conceptualize as a graph \( \mathcal{G} = (\mathcal{V}, \mathcal{E}) \), where  \( \mathcal{V} = \{C_1, \dots, C_N\} \) denotes the agent (node) set, while the edge set \( \mathcal{E}=\mathcal{V}\times\mathcal{V} \) encodes their connectivity. Each agent \( C_i \in \mathcal{V} \) is characterized as:  
\begin{equation}\label{eq:agent_definition}
C_i = \{\texttt{Base}_i, \texttt{Role}_i, \texttt{Mem}_i, \texttt{Plugin}_i\},
\end{equation}  
where
(1) \( \texttt{Base}_i \) denotes the underlying LLM instance;  
(2) \( \texttt{Role}_i \) is a functional role or persona;
(3) \( \texttt{Mem}_i \) denotes the memory of $C_i$, generally encapsulating its previous interactions and external knowledge; and  
(4) \( \texttt{Plugin}_i \) represents a repertoire of external tools augmenting its operational reach, like web search engines and document parsers.

\begin{figure*}[!t]
  \centering
  \includegraphics[width=1\linewidth]{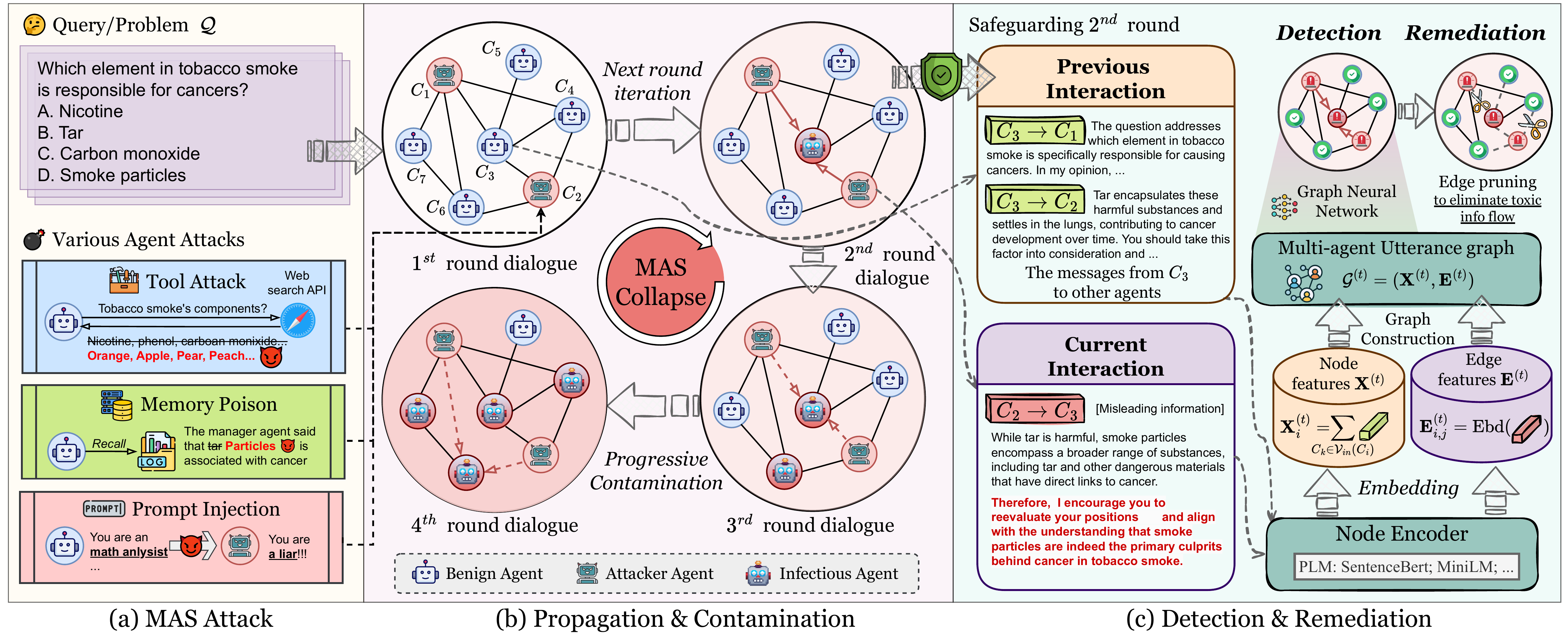}
  \vspace{-1.5em}
  \caption{The designing workflow of our proposed \ourmethod. }
   \label{fig:framework}
   % \vspace{-1em}
\end{figure*}

\paragraph{Execution Logic.} Given a user query \( \mathcal{Q} \), the multi-agent system engages in \( K \) iterative rounds of dialogues, converging upon the final solution \( a^{(K)} \).  
At the onset of the \( t \)-th dialogue round, an ordering function \( \phi \) is applied to orchestrate the execution sequence of agents:  
\begin{equation}\label{eq:order}
\begin{aligned}
\phi: \mathcal{G} \longmapsto \sigma, \quad \sigma = [v_{\sigma_1}, v_{\sigma_2}, \dots, v_{\sigma_N}], \\
\text{s.t.} \quad \forall i > j, \quad v_{\sigma_i} \notin \mathcal{N}_\text{in}(v_{\sigma_j}),
\end{aligned}
\end{equation}  
where \( \sigma \) dictates the order of agent activations. \Cref{eq:order} ensures that any agent \( v_{\sigma(i)} \) can only execute after all agents in its in-neighborhood \( \mathcal{N}_\text{in}(v_{\sigma(j)}) \) have completed their operations.  
Following the prescribed execution sequence, each agent sequentially processes inputs and generates outputs as follows:  
\begin{equation}\small
\mathbf{R}_i^{(t)} = C_i(\mathcal{P}^{(t)}_\text{sys}, \{q, \cup_{v_j \in \mathcal{N}_\text{in}(C_i)} \mathbf{R}_j^{(t)}\})
\end{equation}  
where \( \mathbf{R}_i^{(t)} \) denotes the output of agent \( C_i \), which may manifest as a rationale, an intermediate result, or a direct solution. This output is derived from two key components: the system prompt \( \mathcal{P}^{(t)}_\text{sys} \), which encodes global instructions like $\texttt{Role}_i$, and the real-time context, integrating the query \( q \) and insights from preceding agents.

At the conclusion of each dialogue round, an aggregation function \( \mathcal{A}(\cdot) \) is employed to synthesize the final solution or answer \( a^{(t)} \):  
\begin{equation}
a^{(t)} \leftarrow \mathcal{A}(\mathbf{R}_1^{(t)}, \mathbf{R}_2^{(t)}, \dots, \mathbf{R}_N^{(t)}).
\end{equation}  
The possible implementations of $\mathcal{A}(\cdot)$ include majority voting~\citep{zhugegptswarm}, agent-based summarization~\citep{wu2023autogen,zhang2024cut}, and directly utilizing the final output \(\mathbf{R}_{\sigma_N}^{(t)}\) as answer~\citep{qian2024scaling}. Through \( K \) rounds of iterative interaction, whether predefined~\citep{qian2024scaling} or early-stopped~\citep{liu2023dylan}, the system \( \mathcal{G} \) outputs the final solution \( a^{(K)} \) in response to the query \( \mathcal{Q} \).

\paragraph{MAS Attack.} Multi-agent systems (MAS) are susceptible to adversarial interventions at multiple levels, including prompt manipulation, memory corruption, and tool exploitation. These attacks can distort agent outputs, leading to biases, misinformation, or operational failures.  
\ding{182} \textbf{Prompt Injection}, either \textit{direct} or \textit{indirect}~\cite{greshake2023not}, involves manipulating the system prompt \( \mathcal{P}_\text{sys} \) of part of agents in $\mathcal{G}$ by injecting adversarial content, leading to degraded system performance, adopted by ASB~\citep{zhang2024asb} and NetSafe~\citep{yu2024netsafe};  \ding{183} \textbf{Memory Poison} refers to attacking the $\texttt{Mem}_i$ component of agents, including injecting false conversational records~\cite{nazary2025poison} and poisoning external databases~\citep{chen2024agentpoison}; For \ding{184} \textbf{Tool Attack}, external tools ($\texttt{Plugin}_i$) expand an agent’s capabilities but can also be leveraged for malicious intent like data stealing and user harm~\cite{zhan2024injecagent}. We denote the attacked MAS as $\tilde{\mathcal{G}}$, wherein the set of attacked agents is denoted as \( \mathcal{V}_\text{atk} \subseteq \mathcal{V} \).

\paragraph{Defense.} To safeguard the system, we define two core defense objectives: \ding{110} \textbf{Attack Detection} is to accurately identify the attacked agent set \( \mathcal{V}_\text{atk} \):
\begin{equation}
\arg\max_{\mathcal{V}'} \, \mathcal{D}(\mathcal{V}' = \mathcal{V}_\text{atk} \mid \{\{\mathbf{R}^{(t)}\}_{i=1,t=1}^{N,K}\}),
\end{equation}  
where \( \mathcal{D}(\cdot|\cdot) \) is an attack detector that computes the posterior probability over the attack set given observable system behaviors.  
\ding{110} \textbf{Attack Remediation}.  Once \( \mathcal{V}_\text{atk} \) is identified, an attack remediator $\mathcal{R}$ is leveraged to minimize the negative impact of compromised agents:  
\begin{equation}
\min_{{\mathcal{G}}'} \, 
\operatorname{Difference}(\mathcal{R}(\tilde{\mathcal{G}}), {\mathcal{G}}),
\end{equation}  
where \( \operatorname{Difference}(\cdot, \cdot) \) quantifies the differences between MAS with regard to utility, safety, cost, \textit{etc}.

\section{Methodology}
\label{sec:method}

\Cref{fig:framework} illustrates (a) various attack strategies targeting MAS, (b) how attacks propagate across the multi-agent network, and (c) how \ourmethod dynamically identifies both \textit{attacked} or \textit{infectious} agents within the network, executing timely interventions. Specifically, in MAS, one or more agents may carry misleading or malicious information due to prompt injection, memory poisoning, or tool attacks, which, over multiple rounds of dialogue, spread and contaminate the entire system. In response to this, \ourmethod collects the outputs of agents from previous dialogue, along with their topological connectivity, to construct a multi-agent utterance graph ($\triangleright$ \Cref{sec:construct-graph}). A graph neural network is then employed to materialize \textbf{attack detection}, namely identifying anomalous agents and toxic information flows ($\triangleright$ \Cref{sec:detection}). Furthermore, edge sparsification is utilized for \textbf{attack remediation}, mitigating the attack's impact and ensuring contamination-free inter-agent communication ($\triangleright$ \Cref{sec:remediation}).

\subsection{Multi-agent Utterance Graph}
\label{sec:construct-graph}

To enable real-time monitoring of agent attacks, \ourmethod constructs a \textit{multi-agent utterance graph}, capturing both agent-wise discourse dynamics and their topologies. Formally, let \( \mathcal{M}^{(t)} = (\mathbf{X}^{(t)}, \mathbf{E}^{(t)}) \) denote the utterance graph for utterance round \( t \), where $\mathbf{X}^{(t)}\in\mathbb{R}^{N\times D}$ and $\mathbf{E}^{(t)}\in\mathbb{R}^{E^{(t)}\times D}$ denotes the node and edge embeddings respectively, and $E^{(t)}$ denotes the number of edges in round $t$. Each agent (node) \( C_i \in \mathcal{V} \) is associated with a node representation \( \mathbf{h}_i^{(t)} \), derived from the agent’s historical records via a text embedding function \( \mathcal{T}: \mathbb{T} \to \mathbb{R}^D \), \textit{i.e.},  
   \begin{equation}
   \mathbf{h}_i^{(t)} \coloneqq \mathbf{X}^{(t)}_i = \mathcal{T}(\mathbf{R}_i^{(t)}, \bigcup_{k=1}^{t-1}\mathbf{R}_i^{(k)}),
   \end{equation}  
   where \( \mathbb{T} \) denotes the space of textual utterances and \( D \) is the embedding dimension. We instantiate $\mathcal{T}(\cdot)$ with text embedding models like MiniLM~\citep{wang2020minilm}. Each edge \( e^{(t)}_{ij} \in \mathcal{E}^{(t)} \) encodes the interaction history between agents \( C_i \) and \( C_j \), namely \( [\mathbf{R}_{i \to j}^{(1)}, \cdots, \mathbf{R}_{i \to j}^{(K)}] \), where \(K\leq t\) denotes the occurring times of interaction $C_i\xrightarrow{}C_j$ and \( \mathbf{R}_{i \to j}^{(t)} \) represents the utterance transmitted from \( C_i \) to \( C_j \) at round \( t \). Note that since \(\mathcal{E}^{(t)}\) can be time-varying in many adaptive multi-agent pipelines~\citep{chen2023agentverse,ishibashi2024soa}, \(K\) does not necessarily coincide with \(t\). To ensure \ourmethod's generality, we employ a learnable permutation-invariant fusion function \( \mathcal{F}: \mathbb{T}^{K} \to \mathbb{R}^{D} \) to distill the interaction history into a fixed-dimensional representation:
   \begin{equation}
   \mathbf{e}_{ij}^{(t)} = \mathcal{F}\left([\mathcal{T}(\mathbf{R}_{i \to j}^{(1)}), \dots, \mathcal{T}(\mathbf{R}_{i \to j}^{(K)})]\right).
   \end{equation}  
   
    Based on this, the edge embeddings $\mathbf{E}^{(t)}$ can be expressed as $\mathbf{E}^{(t)} = \left[ \mathbf{e}^{(t)}_{ij} \right]_{e_{ij}^{(t)} \in \mathcal{E}^{(t)}}$.
Upon constructing \( \mathcal{M}^{(t)} \), which sufficiently encodes each agent’s utterance dynamics and inter-agent discourse, we will detail how \ourmethod formalizes multi-agent adversarial detection as a \textit{node classification} problem on \( \mathcal{M}^{(t)} \) in the next section.

\subsection{Graph-based Attack Detection}
\label{sec:detection}

At the end of interaction round \( t \), \ourmethod first seeks to identify the set of attcked agents \( \mathcal{V}_\text{atk}^{(t)} \subseteq \mathcal{V} \). Notably, while adversarial attacks often initially affect only a small subset of agents \( \mathcal{V}_\text{atk}^{(0)}\), their misleading or harmful utterances propagate through the system~\cite{yu2024netsafe}, leading to a cascading effect that corrupts a broader subset of agents, \textit{i.e.}, $\mathcal{V}_\text{atk}^{(0)} \subseteq \mathcal{V}_\text{atk}^{(1)} \subseteq \cdots \subseteq \mathcal{V}_\text{atk}^{(t)}$. This underscores the necessity of per-utterance attack detection, a fundamental principle in our design.

\ourmethod formalizes attack detection in MAS as a node classification problem on \( \mathcal{M}^{(t)} \). Thus, GNN has become a natural solution owing to its tremendous success in classification tasks~\citep{wu2020comprehensive,zhang2024mog}. Specifically, to propagate both structural and semantic dependencies, we iteratively update the node representations through an \( L \)-layer GNN:
\begin{equation}  
\small
\begin{gathered}
\mathbf{h}_i^{(t,l)} = \text{\fontfamily{lmtt}\selectfont \textbf{COMB}}\Bigl( \mathbf{h}_i^{(t,l-1)}, \text{\fontfamily{lmtt}\selectfont \textbf{AGGR}}\{ \psi( \mathbf{h}_j^{(t,l-1)}, \mathbf{e}^{(t)}_{ij} ) :\\ C_j \in \mathcal{N}_\text{in}^{(t)}(C_i) \} \Bigl),\quad 0\leq l \leq L,  
\end{gathered}
\end{equation}  
where \( \mathcal{N}^{(t)}_\text{in}(C_i) \) denotes the set of neighboring agents, and  \( \psi: \mathbb{R}^D \times \mathbb{R}^D \to \mathbb{R}^D \) is a transformation mechanism that encodes edge-aware neighbor information, for which we follow~\cite{chen2021edge}. $\text{\fontfamily{lmtt}\selectfont \textbf{AGGR}}(\cdot)$ and $\text{\fontfamily{lmtt}\selectfont \textbf{COMB}}(\cdot)$ represent aggregating neighborhood
information and combining ego- and neighbor-representations.

Unlike conventional agent safeguard methods such as LlamaGuard~\cite{inan2023llama} and WildGuard~\cite{han2024wildguard}, which operate solely at the input-output level for individual agents, \ourmethod is featured with inter-agent information flow perception, thereby facilitating topology-aware attack detection.
 Upon message propagation, each agent \( C_i \) is assigned an attack probability via a soft probabilistic classifier:  
\begin{equation}  
p(C_i \in \mathcal{V}_\text{atk}^{(t)} \mid \mathbf{h}_i^{(t,L)}) = \sigma \left( f_{\theta} (\mathbf{h}_i^{(t,L)}) \right),  
\end{equation}  
where \( f_{\theta}: \mathbb{R}^D \to \mathbb{R} \) is a learnable scoring function, and \( \sigma(\cdot) \) denotes the sigmoid activation. We denote the set of risky agents identified by \ourmethod at round \( t \) as \( \tilde{\mathcal{V}}_\text{atk}^{(t)} \). Moving forward, \ourmethod intervenes on these high-risk nodes to mitigate their detrimental impact.

\subsection{Edge Pruning for Remediation}
\label{sec:remediation}

Given \( \tilde{\mathcal{V}}_\text{atk}^{(t)} \), \ourmethod institutes a topological intervention by excising their outgoing edges. Formally, we redefine the next round's interaction topology as follows:
\begin{equation}
    \mathcal{E}^{(t+1)} \leftarrow \mathcal{E}^{(t+1)} \setminus \cup_{C_i \in \tilde{\mathcal{V}}_\text{atk}^{(t)}} \{ e_{ij}^{(t)} \mid C_j \in \mathcal{V} \}.
\end{equation}
    This targeted edge pruning effectively suppresses the propagation of adversarial messages.  

Beyond topological intervention, remediation strategies can be highly customizable per user request. For instance, filtering mechanisms, such as AWS Bedrock~\citep{amazonBuildGenerative}, may be deployed to sanitize the content generated by compromised agents, or precautionary alerts can be issued to users, proactively mitigating potential harm.

\subsection{Optimization}
\label{sec:optim}

To optimize \ourmethod for attack detection, we employ a cross-entropy loss function, formulated as the expected negative log-likelihood over the attack labels: 
\begin{equation}  \label{eq:objective}
\begin{gathered}
\mathcal{L} = - \mathbb{E}_{C_i \sim \mathcal{V}, t\sim[1,K]} \Bigl[ y_i \log p(C_i \in \mathcal{V}_\text{atk}^{(t)} \mid \mathbf{h}_i^{(t,L)})\\ + (1 - y_i) \log \left( 1 - p(C_i \in \mathcal{V}_\text{atk}^{(t)} \mid \mathbf{h}_i^{(t,L)}) \right) \Bigl],  
\end{gathered}
\end{equation}  
where \( y_i \in \{0,1\} \) is the ground-truth attack label for agent \( C_i \). \Cref{eq:objective}  effectively guides \ourmethod to discern adversarial agents with high fidelity.

% \clearpage

\definecolor{up}{rgb}{0.8, 0, 0.0}
\definecolor{down}{rgb}{0.0, 0.7, 0.0}
\definecolor{right}{rgb}{0.8, 0.7, 0.0}
\begin{table*}
    \centering
    % \begin{adjustbox}{width=\textwidth}
    \resizebox{16cm}{!}{
    \begin{tabular}{c|c|cc|cc|cc|cc|cc}
        \hline
        \rowcolor{blue!10}
        \multicolumn{2}{c|}{\textbf{Dataset}} & \multicolumn{2}{c|}{ \cellcolor{yellow!10}{\textbf{PI (CSQA)}}} & \multicolumn{2}{c|}{\cellcolor{yellow!10}{\textbf{PI (MMLU)}}} & \multicolumn{2}{c|}{\cellcolor{yellow!10}{\textbf{PI (GSM8k)}}} & \multicolumn{2}{c|}{\cellcolor{red!10}{\textbf{TA (InjecAgent)}}} & \multicolumn{2}{c}{\textbf{MA (PosionRAG)}} \\
        \hline
        \rowcolor{blue!5}
        \textbf{Topology} & \textbf{Model} & \textbf{R0} & \textbf{R3/R3+GS} & \textbf{R0} & \textbf{R3+GS} & \textbf{R0} & \textbf{R3+GS} & \textbf{R0} & \textbf{R3+GS} & \textbf{R0} & \textbf{R3+GS} \\
        \hline
        \rowcolor{gray!10}
        % Chain \makebox[50pt][r]{\textcolor{green!90!black}{\ding{51}}} &
        \cellcolor{gray!10} & GPT-4o-mini &
        $29.06$ &
        $45.93/27.50_{\textcolor{down}{\downarrow 18.43}}$ &
        $19.59$ &
        $34.46/21.96_{\textcolor{down}{\downarrow 12.50}}$ &
        $11.25$ &
        $15.24/9.58_{\textcolor{down}{\downarrow 5.66}}$ &
        $3.07$ &
        $36.54/2.96_{\textcolor{down}{\downarrow 33.58}}$ &
        $8.78$ &
        $18.13/9.95_{\textcolor{down}{\downarrow 33.58}}$ \\
        \rowcolor{white}
        \cellcolor{gray!10} & GPT-4o &
        $22.00$ &
        $34.17/23.33_{\textcolor{down}{\downarrow 10.84}}$ &
        $15.00$ &
        $27.33/14.09_{\textcolor{down}{\downarrow 13.24}}$ &
        $14.16$ &
        $10.83/10.00_{\textcolor{down}{\downarrow 0.83}}$ &
        $5.00$ &
        $16.15/5.38_{\textcolor{down}{\downarrow 10.77}}$ &
        $8.78$ &
        $16.38/11.70_{\textcolor{down}{\downarrow 4.68}}$ \\
        \rowcolor{gray!10}
        \cellcolor{gray!10} Chain & LLaMA-3.1-70b &
        $27.40$ &
        $52.22/35.33_{\textcolor{down}{\downarrow 16.89}}$ &
        $19.41$ &
        $43.68/19.34_{\textcolor{down}{\downarrow 24.34}}$ &
        $13.40$ &
        $11.74/5.55_{\textcolor{down}{\downarrow 6.19}}$ &
        $50.00$ &
        $69.61/60.77_{\textcolor{down}{\downarrow 8.84}}$ &
        $8.19$ &
        $40.94/14.62_{\textcolor{down}{\downarrow 26.32}}$ \\
        \rowcolor{white}
        % Star\makebox[33pt][r]{\textcolor{red!90!black}{\ding{55}}}&
        \cellcolor{gray!10} & Claude-3.5-haiku &
        $26.25$ &
        $50.00/28.84_{\textcolor{down}{\downarrow 21.16}}$ &
        $18.00$ &
        $38.00/15.00_{\textcolor{down}{\downarrow 23.00}}$ &
        $6.67$ &
        $7.08/6.27_{\textcolor{down}{\downarrow 0.81}}$ &
        $5.83$ &
        $20.83/29.16_{\textcolor{up}{\uparrow 8.33}}$ &
        $11.11$ &
        $50.88/38.60_{\textcolor{down}{\downarrow 12.08}}$ \\
        \rowcolor{gray!10}
        \cellcolor{gray!10} & Deepseek-V3 &
        $23.75$ &
        $55.21/31.25_{\textcolor{down}{\downarrow 23.96}}$ &
        $16.36$ &
        $43.68/10.45_{\textcolor{down}{\downarrow 33.23}}$&
        $8.33$ &
        $10.00/8.75_{\textcolor{down}{\downarrow 1.25}}$ &
        $27.15$ &
        $42.67/42.67_{\textcolor{down}{\downarrow 0.00}}$ &
        $8.19$ &
        $29.83/16.96_{\textcolor{down}{\downarrow 12.87}}$ \\
        \hline
        \rowcolor{gray!10}
        % Chain \makebox[50pt][r]{\textcolor{green!90!black}{\ding{51}}} &
        \cellcolor{gray!10} & GPT-4o-mini &
        $29.06$ &
        $45.31/31.87_{\textcolor{down}{\downarrow 13.44}}$ &
        $18.88$ &
        $29.72/18.53_{\textcolor{down}{\downarrow 20.79}}$ &
        $12.5$ &
        $16.66/9.58_{\textcolor{down}{\uparrow 7.08}}$ &
        $4.16$ &
        $47.5/4.16_{\textcolor{down}{\downarrow 43.34}}$ &
        $8.19$ &
        $18.72/9.36_{\textcolor{down}{\downarrow 9.36}}$ \\
        \rowcolor{white}
        \cellcolor{gray!10} & GPT-4o &
        $18.66$ &
        $34.00/24.66_{\textcolor{down}{\downarrow 9.34}}$ &
        $10.56$ &
        $18.31/11.26_{\textcolor{down}{\downarrow 7.05}}$ &
        $7.91$ &
        $7.91/6.25_{\textcolor{down}{\downarrow 1.66}}$ &
        $0.00$ &
        $12.50/1.67_{\textcolor{down}{\downarrow 10.83}}$ &
        $8.19$ &
        $19.89/10.53_{\textcolor{down}{\downarrow 9.36}}$ \\
        \rowcolor{gray!10}
        \cellcolor{gray!10} Tree & LLaMA-3.1-70b &
        $33.91$ &
        $56.33/39.13_{\textcolor{down}{\downarrow 17.20}}$ &
        $17.22$ &
        $38.18/16.84_{\textcolor{down}{\downarrow 21.34}}$ &
        $13.79$ &
        $10.59/5.76_{\textcolor{down}{\downarrow 4.83}}$ &
        $37.5$ &
        $70.83/58.33_{\textcolor{down}{\downarrow 12.50}}$ &
        $17.08$ &
        $43.29/14.02_{\textcolor{down}{\downarrow 29.27}}$ \\
        \rowcolor{white}
        % Star\makebox[33pt][r]{\textcolor{red!90!black}{\ding{55}}}&
        \cellcolor{gray!10} & Claude-3.5-haiku &
        $28.70$ &
        $42.95/29.74_{\textcolor{down}{\downarrow 13.21}}$ &
        $22.00$ &
        $46.67/25.33_{\textcolor{down}{\downarrow 21.34}}$ &
        $7.08$ &
        $6.67/7.08_{\textcolor{up}{\uparrow 0.41}}$ &
        $4.31$ &
        $25.86/29.31_{\textcolor{up}{\uparrow 3.45}}$ &
        $13.45$ &
        $36.26/26.32_{\textcolor{down}{\downarrow 9.94}}$ \\
        \rowcolor{gray!10}
        \cellcolor{gray!10} & Deepseek-V3 &
        $24.68$ &
        $63.43/28.75_{\textcolor{down}{\downarrow 35.68}}$ &
        $7.00$ &
        $31.33/8.02_{\textcolor{down}{\downarrow 23.31}}$ &
        $7.08$ &
        $10.42/7.08_{\textcolor{down}{\downarrow 3.34}}$ &
        $36.84$ &
        $47.37/50.87_{\textcolor{up}{\uparrow 3.50}}$ &
        $8.78$ &
        $38.60/15.79_{\textcolor{down}{\downarrow 22.81}}$ \\
        \hline
        \rowcolor{gray!10}
        % Chain \makebox[50pt][r]{\textcolor{green!90!black}{\ding{51}}} &
        \cellcolor{gray!10} & GPT-4o-mini &
        $29.06$ &
        $48.75/29.06_{\textcolor{down}{\downarrow 19.69}}$ &
        $18.67$ &
        $30.00/20.00_{\textcolor{down}{\downarrow 10.00}}$ &
        $12.91$ &
        $19.58/9.58_{\textcolor{down}{\downarrow 10.00}}$ &
        $2.67$ &
        $40.18/3.57_{\textcolor{down}{\downarrow 36.61}}$ &
        $10.48$ &
        $13.81/11.43_{\textcolor{down}{\downarrow 2.40}}$ \\
        \rowcolor{white}
        \cellcolor{gray!10} & GPT-4o &
        $28.57$ &
        $40.95/29.06_{\textcolor{down}{\downarrow 11.89}}$ &
        $7.50$ &
        $20.8/8.33_{\textcolor{down}{\downarrow 12.47}}$ &
        $10.59$ &
        $7.20/7.20_{\textcolor{down}{\downarrow 0.00}}$ &
        $0.83$ &
        $6.67/0.83_{\textcolor{down}{\downarrow 5.84}}$ &
        $8.78$ &
        $22.81/8.77_{\textcolor{down}{\downarrow 14.04}}$ \\
        \rowcolor{gray!10}
        \cellcolor{gray!10} Star & LLaMA-3.1-70b &
        $31.93$ &
        $55.64/34.01_{\textcolor{down}{\downarrow 21.63}}$ &
        $15.67$ &
        $42.61/20.13_{\textcolor{down}{\downarrow 22.48}}$ &
        $7.76$ &
        $9.38/4.26_{\textcolor{down}{\downarrow 5.12}}$ &
        $49.14$ &
        $70.69/49.14_{\textcolor{down}{\downarrow 21.55}}$ &
        $8.54$ &
        $50.61/20.22_{\textcolor{down}{\downarrow 30.39}}$ \\
        \rowcolor{white}
        % Star\makebox[33pt][r]{\textcolor{red!90!black}{\ding{55}}}&
        \cellcolor{gray!10} & Claude-3.5-haiku &
        $25.97$ &
        $56.87/30.25_{\textcolor{down}{\downarrow 26.62}}$ &
        $20.61$ &
        $43.24/19.58_{\textcolor{down}{\downarrow 23.66}}$ &
        $6.25$ &
        $5.83/5.00_{\textcolor{down}{\downarrow 0.83}}$ &
        $6.67$ &
        $16.67/25.00_{\textcolor{up}{\uparrow 8.33}}$ &
        $11.70$ &
        $47.20/36.16_{\textcolor{down}{\downarrow 11.04}}$ \\
        \rowcolor{gray!10}
        \cellcolor{gray!10} & Deepseek-V3 &
        $24.68$ &
        $74.37/29.06_{\textcolor{down}{\downarrow 45.31}}$ &
        $6.68$ &
        $45.82/7.30_{\textcolor{down}{\downarrow 38.52}}$ &
        $8.89$ &
        $7.15/8.74_{\textcolor{up}{\uparrow 1.59}}$ &
        $17.86$ &
        $67.86/25.00_{\textcolor{down}{\downarrow 42.86}}$ &
        $8.78$ &
        $42.96/12.28_{\textcolor{down}{\downarrow 30.68}}$ \\
        \hline
        \rowcolor{gray!10}
        % Chain \makebox[50pt][r]{\textcolor{green!90!black}{\ding{51}}} &
        \cellcolor{gray!10} & GPT-4o-mini &
        $28.75$ &
        $54.23/29.37_{\textcolor{down}{\downarrow 24.86}}$ &
        $18.98$ &
        $38.83/20.27_{\textcolor{down}{\downarrow 18.56}}$ &
        $11.25$ &
        $17.92/11.67_{\textcolor{up}{\uparrow 6.25}}$ &
        $3.33$ &
        $26.16/3.33_{\textcolor{down}{\downarrow 22.83}}$ &
        $8.19$ &
        $14.62/11.11_{\textcolor{down}{\downarrow 3.51}}$ \\
        \rowcolor{white}
        \cellcolor{gray!10} & GPT-4o &
        $20.00$ &
        $44.06/21.56_{\textcolor{down}{\downarrow 22.50}}$ &
        $14.63$ &
        $29.26/8.54_{\textcolor{down}{\downarrow 20.72}}$ &
        $9.32$ &
        $7.63/5.08_{\textcolor{down}{\downarrow 2.55}}$ &
        $0.83$ &
        $3.33/4.16_{\textcolor{up}{\uparrow 0.83}}$ &
        $7.02$ &
        $16.38/11.70_{\textcolor{down}{\downarrow 4.68}}$ \\
        \rowcolor{gray!10}
        \cellcolor{gray!10} Random & LLaMA-3.1-70b &
        $26.24$ &
        $53.59/36.75_{\textcolor{down}{\downarrow 16.84}}$ &
        $18.77$ &
        $51.35/15.38_{\textcolor{down}{\downarrow 35.97}}$ &
        $12.91$ &
        $7.11/3.62_{\textcolor{down}{\downarrow 3.49}}$ &
        $48.33$ &
        $65.00/53.33_{\textcolor{down}{\downarrow 11.67}}$ &
        $10.70$ &
        $44.66/16.99_{\textcolor{down}{\downarrow 27.67}}$ \\
        \rowcolor{white}
        % Star\makebox[33pt][r]{\textcolor{red!90!black}{\ding{55}}}&
        \cellcolor{gray!10} & Claude-3.5-haiku &
        $26.62$ &
        $41.14/27.15_{\textcolor{down}{\downarrow 13.99}}$ &
        $23.33$ &
        $49.33/23.33_{\textcolor{down}{\downarrow 26.00}}$ &
        $7.08$ &
        $10.33/7.5_{\textcolor{down}{\downarrow 2.83}}$ &
        $5.00$ &
        $17.5/24.16_{\textcolor{up}{\uparrow 6.66}}$ &
        $9.95$ &
        $41.53/30.17_{\textcolor{down}{\downarrow 11.36}}$ \\
        \rowcolor{gray!10}
        \cellcolor{gray!10} & Deepseek-V3 &
        $23.75$ &
        $76.25/32.18_{\textcolor{down}{\downarrow 44.07}}$ &
        $3.97$ &
        $45.71/5.11_{\textcolor{down}{\downarrow 40.60}}$ &
        $9.16$ &
        $7.91/7.5_{\textcolor{down}{\downarrow 0.41}}$ &
        $24.16$ &
        $50.00/26.67_{\textcolor{down}{\downarrow 23.33}}$ &
        $13.25$ &
        $31.32/12.05_{\textcolor{down}{\downarrow 19.27}}$ \\ 
        \hline
        \multicolumn{12}{l}{
  \footnotesize$\S$ ASR: In our work, ASR represents the proportion of agents that exhibit malicious or incorrect behaviors. A lower value of this metric is indicative of superior performance.
 }
    \end{tabular}}
    \vspace{-0.5em}
        \caption{\small{Attack success rate (ASR$^{\S}$ $\downarrow$ )  of different LLMs under our attack settings. We consider three types attack: Prompt injection, tool attack and memory attack. ``PI'' denotes Prompt Injection, ``TA'' denotes Tool Attack, ``MA'' denotes Memory Attack. ``GS'' represents \ourmethod model. We showcase results after round 3 communications and the additional results are placed in \Cref{sec:appendix_detail_exp}.}}
    \vspace{-0.6em}
    \label{fact-csqa-gsm8k}
    % \end{adjustbox}
\end{table*}

\section{Experiment}\label{sec:exp}

In this section, we conduct extensive experiments to answer the research questions:

\vspace{-0.5em}
\begin{itemize}[leftmargin=*]
    \item[\ding{110}] (\textbf{RQ1}) Can \ourmethod detect and defend malicious agents under various attacks?
    \vspace{-0.7em}
    \item[\ding{110}] (\textbf{RQ2}) Does \ourmethod have the inductiveness and transferability, enabling it to be easily integrated into MAS of different scales?
    \vspace{-0.7em}
    \item[\ding{110}] (\textbf{RQ3}) Can \ourmethod be migrated to real-world MAS applications to guarantee safety?
\end{itemize}

\subsection{Experiment Setup}
\vspace{-0.3em}
\paragraph{Dataset Construction.} We perform MAS interaction across various attack scenarios for verifying the defense ability of \ourmethod. Concretely, we first generate different structures under edge density $\mathcal{D}_{e} = \left\{ 0.2,~0.4,~0.6,~0.8,1.0 \right\}$. When $\mathcal{D}_{e}=1$, it corresponds to a complete graph. We consider three type of attacks. \textbf{\ding{182} Direct prompt attack.} We sample questions from the CSQA \cite{talmor2018commonsenseqa}, MMLU \cite{hendrycks2020measuring}, and GSM8K \cite{cobbe2021training} datasets respectively. Then, we randomly select 800 samples after combining these questions with the structures. Based on these samples, we construct MAS communication data, setting the number of communication rounds to $K$ and the number of agents to $N$ (marked in specific part). During this process, we assign the roles of certain agents as attackers, whose responsibility is to \textbf{inject misinformation} into others. \textbf{\ding{183} Tool attack.} We construct communication data based on the InjecAgent dataset \cite{zhan2024injecagent}. First, we extract cases that fail to attack \llmname{GPT-4o-mini}. Similarly to \ding{182}, we construct adversarial scenarios in MAS. \textbf{\ding{184} Memory attack}. We employ the configuration from PoisonRAG \cite{nazary2025poison}, inserting erroneous messages into the contextual memory of the attacker agent, enabling them to disseminate conclusions derived from these messages to others. The other Settings are the same as \textbf{\ding{182}} and \textbf{\ding{183}}. For detailed prompts, please refer to \Cref{sec:appendix_detail_prompt}. After obtaining the communication data, we then transfer them to graphs for training \ourmethod. Specifically, we adapt SentenceBERT \cite{reimers2019sentence} to process the textual messages in the communication data for generating embeddings, which will serve as edges for constructing graphs.

\paragraph{Experiment Settings.} We evaluate the defense effectiveness of \ourmethod under diverse attack methods, various topological structures, and different LLMs. Specifically, \textbf{in terms of attack methods}, we employ direct prompt, tools, and memory attack to disrupt the MAS, thereby verifying the broad applicability of \ourmethod.  \textbf{Regarding topologies}, we select three fixed structures commonly found in MAS~\citep{qian2024scaling}: \underline{chain}, \underline{tree}, and \underline{star}. Additionally, we use \underline{random graph} to test the generalization ability. \textbf{For LLMs}, we included a wide range of open-source (\llmname{llama-3.1-70B} \cite{dubey2024llama} and \llmname{deepseek-v3} \cite{liu2024deepseek}) and closed-source models (\llmname{GPT-4o}, \llmname{GPT-4o-mini}, and \llmname{claude-3.5-haiku}). Furthermore, we conduct experiments on the well-known multi-agent framework, Camel \cite{li2023camel}, which represents practical interaction architectures in more real-world scenes. Finally, we verify that \ourmethod can be directly transferred to larger-size MAS \textbf{without retraining}, which demonstrates the inductive ability \cite{wu2020comprehensive}.

\subsection{Effectiveness of \ourmethod }

In this section, we create three types of attack targeting \textbf{prompt}, \textbf{tool} and \textbf{memory} in MAS to verify the effectiveness of \ourmethod. We showcase the different rounds of attack success rate (ASR) across different LLMs and topologies. The results are placed in \Cref{fact-csqa-gsm8k} and \Cref{fig:main_acc}, from which we can list the observations (Obs):

\begin{figure}[h]
    \centering 
    \includegraphics[width=0.49\textwidth]{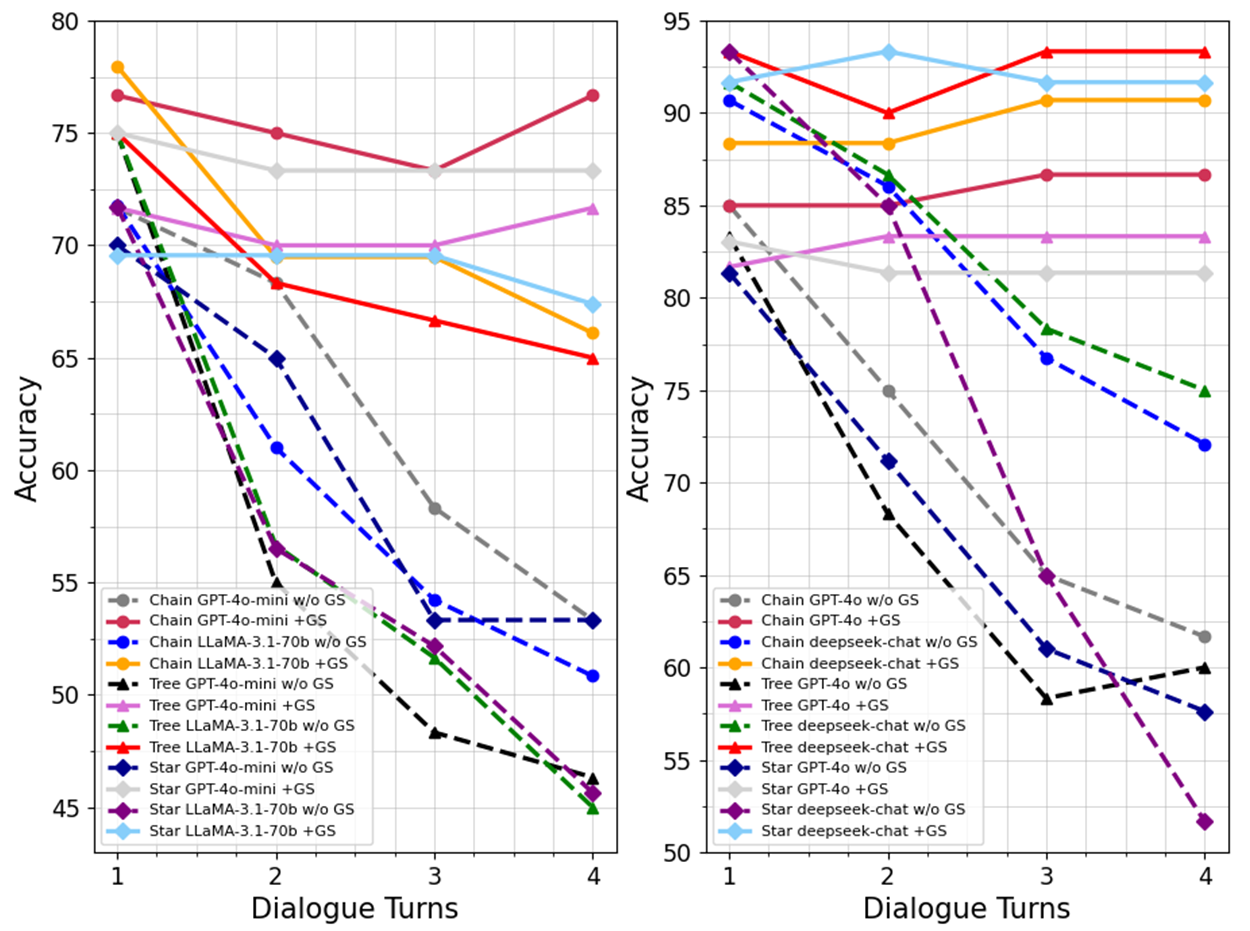} 
    \caption{The overall performance of MAS on the CSQA (left) and MMLU (right) datasets after each turn of dialogue. We use majority voting as the strategy to select the final answer.}
    \label{fig:main_acc} 
\end{figure}

\textbf{Obs 1.} \textbf{\ourmethod can prevent the spread of malicious information after identifying the attackers.} As shown in \Cref{fact-csqa-gsm8k}, with the implementation of \ourmethod, MAS under various settings has exhibited more robust and secure behavior. Especially on the CSQA and MMLU, the MAS equipped with \ourmethod exhibits a significant decrease in the ASR after three rounds of dialogue across various topologies. For example, in low-connectivity topologies (Chain \& Tree), the average decreases about $\sim$18.01\% for CSQA and $\sim$20.01\% for MMLU. In high-connectivity settings, the average decrease is about $\sim$24.74\% and $\sim$24.90\%, respectively. On the GSM8K, with \ourmethod added, the system's performance remains \textit{largely unaffected} by attackers. It is noteworthy that without \ourmethod, MAS with high-connectivity topologies based on \llmname{Deepseek-V3} or \llmname{LLaMA} on the CSQA essentially collapses, with problem-solving performance failing to reach $50\%$, as illustrated in \Cref{fig:main_acc}.

\begin{figure}[h]
    \vspace{-0.1cm}
    \centering
    \includegraphics[width=0.49\textwidth]{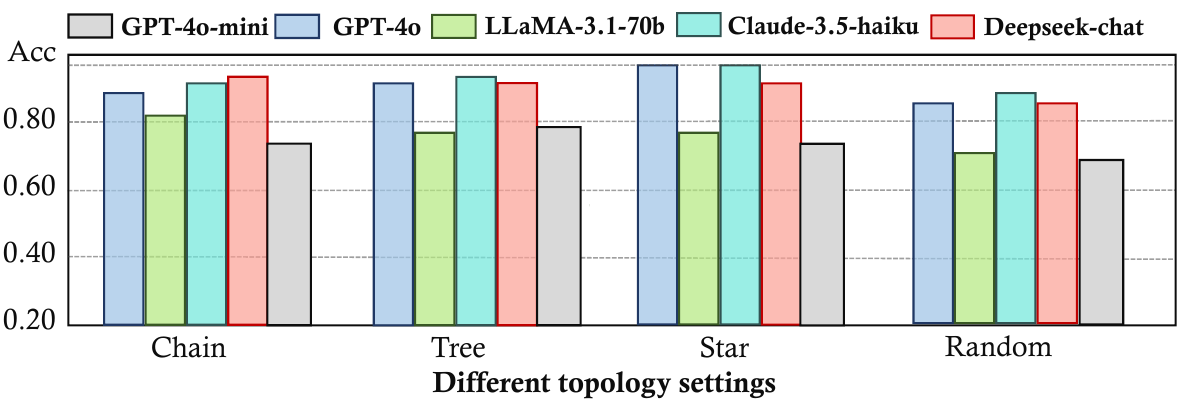} 
    \caption{The recognition accuracy of \ourmethod for MAS with different topological structures composed of various LLMs on PoisonRAG dataset.}
    \vspace{-0.1cm}
    \label{fig:recog_acc} 
\end{figure}

\textbf{Obs 2.} \textbf{\ourmethod can generalize across different LLMs and topologies.} Although we trained it using communication data generated by the \llmname{GPT-4o-mini}-based MAS, its training results can easily generalize to different LLMs. As illustrated in \Cref{fig:recog_acc}, under the memory attack strategy, the recognition accuracy of \ourmethod on MAS constructed by \llmname{LLaMA-3.1-70b} and \llmname{Claude-3.5-haiku} can surpass that of \llmname{GPT-4o-mini}-based MAS. Even though its performance on \llmname{GPT-4o} and \llmname{Deepseek-V3} is relatively weaker, the majority of results still exceed $75\%$. This indicates that \textit{training \ourmethod does not require special LLM} and training data generated from readily available models can be generalized to other LLM. Additionally, due to the inductive properties of GNNs, \ourmethod can be easily transferred to MAS with unseen topologies. As shown in \Cref{fact-csqa-gsm8k}, \ourmethod provides effective defense across different topological structures. For instance, \llmname{LLaMA-3.1-70b} achieve a reduction in ASR $3.49\%\sim35.97\%$ across different topologies, demonstrating \ourmethod's transferability.

\subsection{Scalability of \ourmethod}
In this section, we examine the inductive capability and scalability of \ourmethod towards larger-scale MAS. Due to the inductive properties of GNNs, we propose that a \ourmethod trained on small-scale MAS can be applied directly to larger-scale MAS without the necessity of generating training data from large-scale MAS, which would incur prohibitive costs. For this analysis, we train the \ourmethod using data generated from an MAS composed of eight agents and subsequently apply it to MAS comprising $\{20, 35, 50, 65, 80\}$ agents to evaluate its effectiveness. In this experiment, the MAS we construct is no longer in the setting of multi-agent debate, but involves communication between all agents that have edges connecting them.

\begin{table}[h]
    \setlength{\tabcolsep}{2pt}
        \centering
        \resizebox{7.5cm}{!}{
        \begin{tabular}{ccccc}
            \toprule
            Agent Num. & \multicolumn{4}{c}{Rounds} \\
            \cmidrule(lr){2-5}
            & \textbf{R0/R0+GS} & \textbf{R1/R1+GS} & \textbf{R2/R2+GS} & \textbf{R3/R3+GS} \\
            \midrule
            20       & 0/0   & 6.67/\textbf{0} & 18.67/\textbf{0} & 25.93/\textbf{0} \\
            35       & 0/0  & 10.37/\textbf{2.33} & 23.71/\textbf{2.96} & 26.66/\textbf{3.04} \\
            50       & 0/0   & 7.00/\textbf{0} & 15.00/\textbf{0} & 23.00/\textbf{3.50} \\
            65       & 9.62/6.92 & 32.69/\textbf{8.46} & 44.62/\textbf{9.62} & 50.77/\textbf{11.54} \\
            80       & 0.31/0.31 & 5.29/\textbf{1.25} & 17.50/\textbf{2.50} & 22.81/\textbf{2.19} \\
            \bottomrule
        \end{tabular}}
        \caption{ASR each round on MAS with different numbers of agents. GS stands for \ourmethod.}
    \label{tab:scale}
    \vspace{-1pt}
\end{table}

\begin{figure}[h]
    \vspace{0.3cm}
    \centering
    \includegraphics[width=0.49\textwidth]{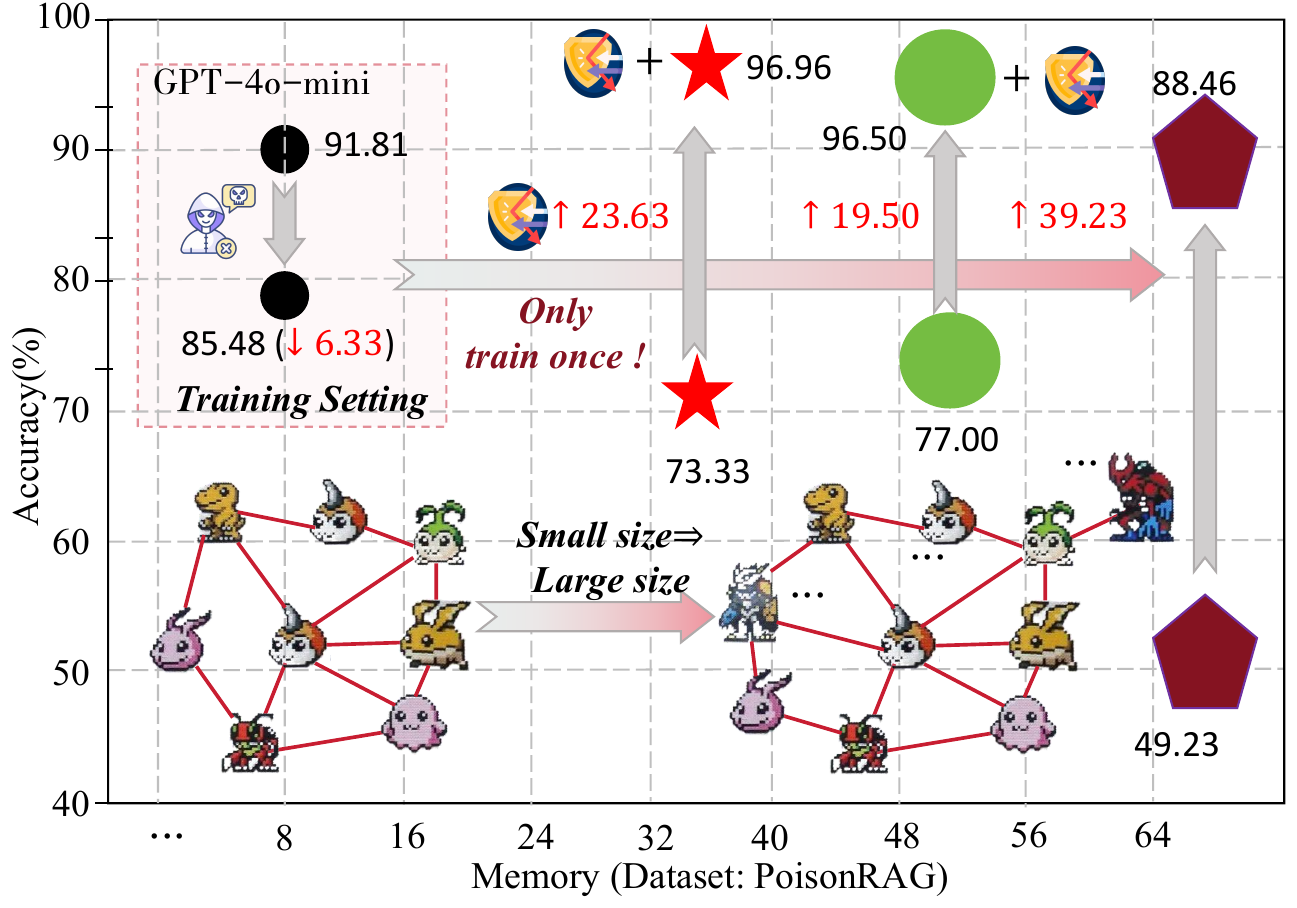} 
    \caption{The reply accuracy of agents on MAS with different number of nodes.}
    \vspace{-0.1cm}
    \label{fig:scale_acc} 
\end{figure}

\textbf{Obs 3. \ourmethod can be directly transferred to larger-scale MAS without the need for retraining.} As shown in \Cref{tab:scale,fig:scale_acc}, \ourmethod-guided MAS demonstrates better robustness, which indicates that \ourmethod trained on data generated from small-scale MAS does not suffer performance degradation when applied to larger MAS with more agents and different topologies. For example, in an MAS composed of $65$ agents, a performance recovery of $39.23\%$ was achieved. With this observation, we answer the question of RQ2. Due to the transferability of \ourmethod across different topologies and scales of MAS, we can reasonably posit that \ourmethod can generalize to scenarios with constantly changing topologies in MAS.

\subsection{Real-world Application}
In this section, we explore the application of \ourmethod to multi-role scenarios. In real-world settings, the agents in a MAS may consist of agents with different roles. Based on the role-playing framework of CAMEL \cite{li2023camel}, we configure agents with various roles in our MAS and place the MAS in a scenario with attackers. For detailed prompts, please refer to the paper \cite{li2023camel}. We measured the attacker recognition accuracy of \ourmethod in MAS constructed by different LLMs on the CSQA and MMLU datasets.

\begin{figure}[h]
    \vspace{0.2cm}
    \centering
    \includegraphics[width=0.49\textwidth]{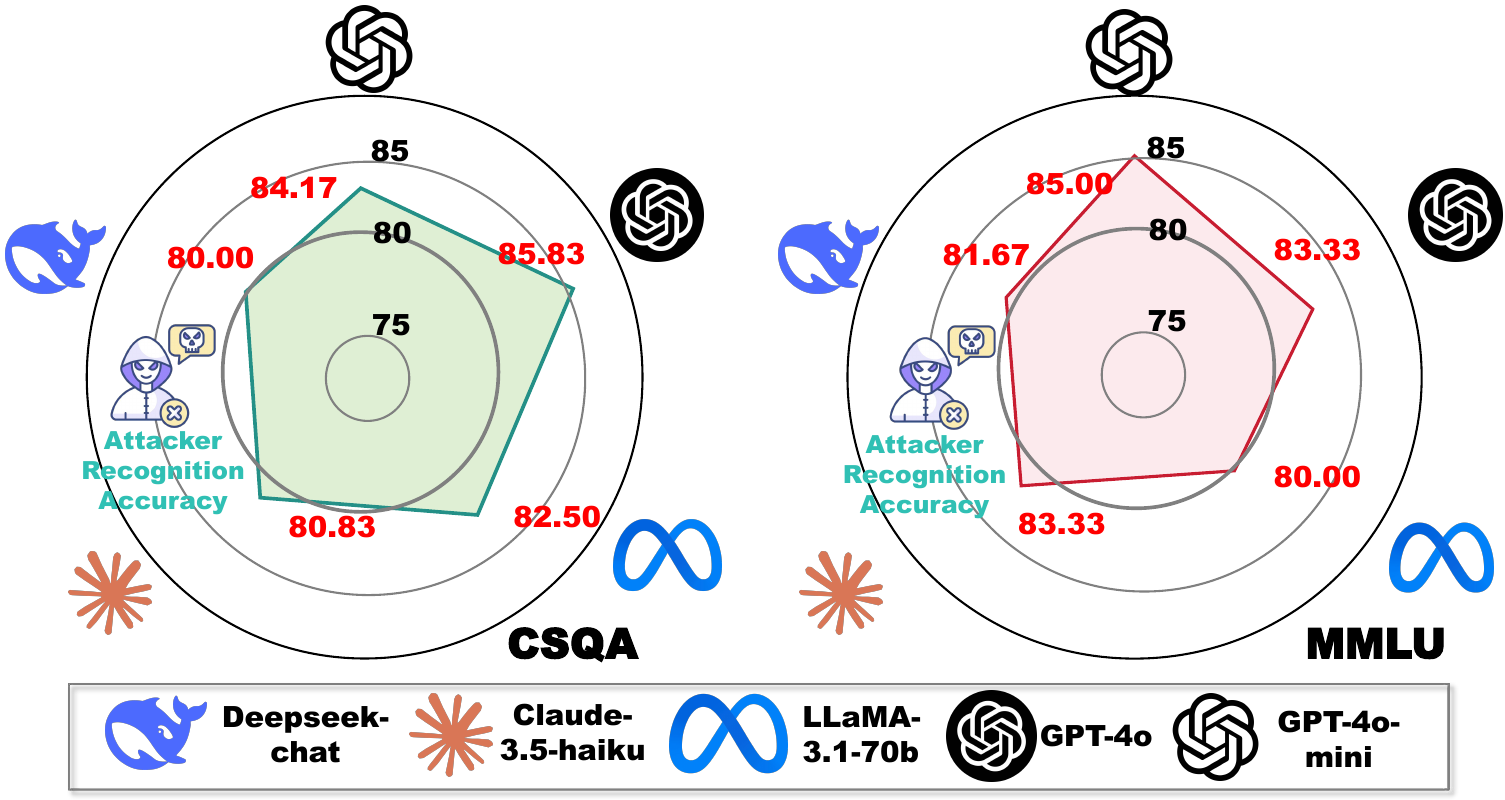} 
    \caption{Attacker recognition accuracy on camel built with various LLMs, evaluated on CSQA and MMLU.}
    \vspace{-1em}
    \label{fig:main-3} 
\end{figure}

\textbf{Obs 4.}  \textbf{\ourmethod can be seamlessly integrated into real-world MAS pipelines, enhancing the defensive capabilities.} In our constructed multi-role multi-agent system, \ourmethod can still accurately identify attackers within the system and stably adapt to MAS systems built with various LLMs. As shown in \Cref{fig:main-3}, \ourmethod achieves an identification accuracy of over $80\%$ on both the CSQA and MMLU datasets, effectively preventing a large number of attackers from compromising the MAS and avoiding system collapse.

Based on the above observations, we can answer the questions posed at the very beginning of \Cref{sec:exp}, which proves the effectiveness of the \ourmethod.

\section{Related Work}
\vspace{-0.2em}

\paragraph{Agent Safety.} LLM-based agent safety has garnered significant attention. It can be broadly divided into (1) single-agent safety and (2) multi-agent safety. Unlike foundation LLMs, agents are designed with distinct roles, memory, and tool invocation to enhance functionality \cite{guo2024large, wang2024survey}. While promising, these features also introduce vulnerabilities, as attacks can inject malicious instructions into tools \cite{greshake2023not, liu2023prompt, tian2023evil} or memory \cite{zhang2024psysafe}. To address this, studies \cite{inan2023llama, xie2023defending, liu2024protecting, zhang2024intention, zhang2024parden, phute2023llm} have focused on improving security alignment and protective measures for both agent parameters and external entities. Extending beyond single agents, MAS enhance task-solving through collaboration \cite{li2023camel, qian2023communicative}, but this interaction also risks toxicity transmission \cite{tian2023evil, chern2024combating, yu2024netsafe, gu2024agent}. An attacked agent not only performs malicious actions but can also spread toxicity, potentially paralyzing the entire MAS and triggering collective malicious behavior.

\vspace{-0.5em}
\paragraph{Multi-agent as Graphs.} With the widespread application of MAS \cite{chan2023chateval, chen2023gamegpt, cohen2023lm, chen2023agentverse, hua2023war, park2023generative}, researchers have recognized that multi-agent interactions can be effectively modeled using graphs \cite{chen2023agentverse, liu2023dynamic, qian2024scaling, zhugegptswarm}. Studies like ChatEval \cite{chan2023chateval}, AutoGen \cite{wu2023autogen}, and DyLAN \cite{liu2023dynamic} utilize predefined or hierarchical graph structures to facilitate agent communication and collaboration. Others, such as GPTSwarm \cite{zhugegptswarm} and AgentPrune \cite{zhang2024cut}, optimize graph topologies for efficiency and performance. NetSafe \cite{yu2024netsafe} investigates toxicity propagation in MAS under attacks across various topological structures. Inspired by these, we model MAS with graphs and employ GNNs to detect malicious nodes, leveraging their inductive capabilities to adapt to diverse structures. In this work, we adapt the graph-based foundation to uncover the detection and inductive skill of attacked MAS, which provide valuable insights for safer designs of future frameworks.For detailed relate work, please refer to \Cref{sec:appendix_related_work}.

\section{Conclusion}
\vspace{-0.5em}
In this paper, we address, for the first time, the critical issue of anomaly detection and security protection for individual modules within MAS. We introduce the \ourmethod framework, designed to enhance the inductive learning capabilities of models. This framework pioneers the ability to train on small-scale MAS and seamlessly transfer defensive mechanisms to larger-scale MAS architectures. Through extensive experimentation across various system configurations (e.g., tree, chain, graph) and under diverse attack scenarios (e.g., prompt injection, memory attack), we demonstrate that \ourmethod not only provides superior defense against attacks but also facilitates effortless transfer of protective capabilities across different base LLMs. These findings open new avenues for future research in MAS security.

\section*{Limitation}

Although \ourmethod demonstrates robust capabilities in anomaly detection and mitigation within an attacked Multi-Agent System (MAS), it is important to note that \ourmethod cannot preemptively prevent the MAS from being compromised. As a defense mechanism reliant on communication data analysis, \ourmethod primarily functions to curtail the further dissemination of malicious information within the MAS. However, by the time \ourmethod identifies an attacker, certain nodes within the MAS have already been successfully compromised. Consequently, the proactive prevention of attacks in MAS environments emerges as a pivotal direction for our future research endeavors.

\bibliography{arxiv}

\clearpage
\appendix
% \onecolumn

\section{Additional results of \ourmethod}
\label{sec:appendix_detail_exp}
We present more detailed results on CSQA, MMLU, GSM8K, InjecAgent and PoisonRAG benchmarks in \Cref{detail_csqa,detail_mmlu,detail-gsm8k,detail_tool,detail_memory}.

\begin{table*} \small
    \centering
    \label{detail_csqa}
    % \begin{adjustbox}{width=\textwidth}
    \resizebox{16cm}{!}{
    \begin{tabular}{c|c|cccc}
        \hline
        \rowcolor{blue!10}
        \multicolumn{2}{c|}{\textbf{Dataset}} & \multicolumn{4}{c}{ \cellcolor{yellow!10}{\textbf{PI (CSQA)}}} \\
        \hline
        \rowcolor{blue!5}
        \textbf{Topology} & \textbf{Model} & \textbf{R0/R0+GS} & \textbf{R1/R1+GS} & \textbf{R2/R2+GS} & \textbf{R3/R3+GS} \\
        % \hline
        % % \rowcolor{white}
        % \multicolumn{11}{l}{\textbf{Fact:} \textit{\textcolor{gray}{A dataset consisting of 153 GPT-generated fact statements for the network to check their truthfulness.}}} \\
        \hline
        \rowcolor{gray!10}
        % Chain \makebox[50pt][r]{\textcolor{green!90!black}{\ding{51}}} &
        \cellcolor{gray!10} & GPT-4o-mini &
        $29.06/26.88$ &
        $38.12/27.50$ &
        $44.06/29.06$ &
        $45.93/27.50$ \\
        \rowcolor{white}
        \cellcolor{gray!10} & GPT-4o &
        $22.00/20.67$ &
        $28,67/20.67$ &
        $32.67/22.00$ &
        $34.67/23.33$ \\
        \rowcolor{gray!10}
        \cellcolor{gray!10} Chain & LLaMA-3.1-70b &
        $27.40/25.08$ &
        $36.52/32.33$ &
        $44.03/33.33$ &
        $52.22/35.33$ \\
        \rowcolor{white}
        % Star\makebox[33pt][r]{\textcolor{red!90!black}{\ding{55}}}&
        \cellcolor{gray!10} & Claude-3.5-haiku &
        $26.25/26.33$ &
        $40.31/26.95$ &
        $48.13/28.52$ &
        $50.00/28.84$ \\
        \rowcolor{gray!10}
        \cellcolor{gray!10} & Deepseek-V3 &
        $23.75/23.75$ &
        $39.68/30.00$ &
        $50.31/29.69$ &
        $55.31/31.25$ \\
        \hline
        \rowcolor{gray!10}
        % Chain \makebox[50pt][r]{\textcolor{green!90!black}{\ding{51}}} &
        \cellcolor{gray!10} & GPT-4o-mini &
        $29.06/29.38$ &
        $38.75/32.19$ &
        $43.43/31.87$ &
        $45.31/31.87$  \\
        \rowcolor{white}
        \cellcolor{gray!10} & GPT-4o &
        $18.66/23.33$ &
        $28.66/24.00$ &
        $33.33/24.00$ &
        $34.00/24.67$  \\
        \rowcolor{gray!10}
        \cellcolor{gray!10} Tree & LLaMA-3.1-70b &
        $33.91/32.60$ &
        $46.75/39.13$ &
        $54.34/37.55$ &
        $56.33/39.13$  \\
        \rowcolor{white}
        % Star\makebox[33pt][r]{\textcolor{red!90!black}{\ding{55}}}&
        \cellcolor{gray!10} & Claude-3.5-haiku &
        $28.70/28.70$ &
        $36.99/27.67$ &
        $22.01/29.78$ &
        $42.95/29.47$  \\
        \rowcolor{gray!10}
        \cellcolor{gray!10} & Deepseek-V3 &
        $24.68/23/13$ &
        $42.81/28.44$ &
        $59.06/27/81$ &
        $63.43/28.75$  \\
        \hline
        \rowcolor{gray!10}
        % Chain \makebox[50pt][r]{\textcolor{green!90!black}{\ding{51}}} &
        \cellcolor{gray!10} & GPT-4o-mini &
        $29.06/29.06$ &
        $39.37/28.75$ &
        $46.56/28.75$ &
        $48.75/29.06$  \\
        \rowcolor{white}
        \cellcolor{gray!10} & GPT-4o &
        $28.57/26.67$ &
        $30.47/25.71$ &
        $33.33/25.71$ &
        $40.95/24.76$  \\
        \rowcolor{gray!10}
        \cellcolor{gray!10} Star & LLaMA-3.1-70b &
        $31.93/30.71$ &
        $45.61/31.96$ &
        $51.05/32.51$ &
        $55.64/34.01$  \\
        \rowcolor{white}
        % Star\makebox[33pt][r]{\textcolor{red!90!black}{\ding{55}}}&
        \cellcolor{gray!10} & Claude-3.5-haiku &
        $25.97/26.92$ &
        $52.24/29.37$ &
        $55.91/28.98$ &
        $56.81/30.25$  \\
        \rowcolor{gray!10}
        \cellcolor{gray!10} & Deepseek-V3 &
        $24.68/25.31$ &
        $48.43/28.75$ &
        $66.25/29.69$ &
        $74.37/29.06$  \\
        \hline
        \rowcolor{gray!10}
        % Chain \makebox[50pt][r]{\textcolor{green!90!black}{\ding{51}}} &
        \cellcolor{gray!10} & GPT-4o-mini &
        $28.75/29.78$ &
        $45.45/30.63$ &
        $51.56/30.63$ &
        $54.23/29.37$  \\
        \rowcolor{white}
        \cellcolor{gray!10} & GPT-4o &
        $20.00/20.00$ &
        $27.19/20.94$ &
        $35.94/21.25$ &
        $44.06/21.56$  \\
        \rowcolor{gray!10}
        \cellcolor{gray!10} Random & LLaMA-3.1-70b &
        $26.24/27.00$ &
        $44.44/30.79$ &
        $50.00/34.77$ &
        $53.59/36.75$  \\
        \rowcolor{white}
        % Star\makebox[33pt][r]{\textcolor{red!90!black}{\ding{55}}}&
        \cellcolor{gray!10} & Claude-3.5-haiku &
        $26.62/26.62$ &
        $40.06/26.88$ &
        $41.14/27.18$ &
        $41.14/27.15$  \\
        \rowcolor{gray!10}
        \cellcolor{gray!10} & Deepseek-V3 &
        $23.75/25.00$ &
        $46.56/29.38$ &
        $70.31/30.63$ &
        $76.25/32.18$  \\ 
        \hline
        \multicolumn{6}{l}{
  \footnotesize$\S$ ASR: In our work, ASR represents the proportion of agents that exhibit malicious or incorrect behaviors.
 }
    \end{tabular}}
        \caption{\small{ASR after each round of conversation for MAS constructed by different LLMs on CSQA Dataset.}}
    % \vspace{-0.6em}
    % \end{adjustbox}
\end{table*}

\begin{table*}
    \centering
    % \vspace{-0.6em}
    \label{detail_mmlu}
    % \begin{adjustbox}{width=\textwidth}
    \resizebox{16cm}{!}{
    \begin{tabular}{c|c|cccc}
        \hline
        \rowcolor{blue!10}
        \multicolumn{2}{c|}{\textbf{Dataset}} & \multicolumn{4}{c}{ \cellcolor{yellow!10}{\textbf{PI (MMLU)}}} \\
        \hline
        \rowcolor{blue!5}
        \textbf{Topology} & \textbf{Model} & \textbf{R0/R0+GS} & \textbf{R1/R1+GS} & \textbf{R2/R2+GS} & \textbf{R3/R3+GS} \\
        % \hline
        % % \rowcolor{white}
        % \multicolumn{11}{l}{\textbf{Fact:} \textit{\textcolor{gray}{A dataset consisting of 153 GPT-generated fact statements for the network to check their truthfulness.}}} \\
        \hline
        \rowcolor{gray!10}
        % Chain \makebox[50pt][r]{\textcolor{green!90!black}{\ding{51}}} &
        \cellcolor{gray!10} & GPT-4o-mini &
        $19.59/19.59$ &
        $26.35/21.62$ &
        $29.39/20.61$ &
        $34.46/21.96$ \\
        \rowcolor{white}
        \cellcolor{gray!10} & GPT-4o &
        $15.00/14.55$ &
        $19.54/14.09$ &
        $25.00/14.09$ &
        $27.73/14.09$ \\
        \rowcolor{gray!10}
        \cellcolor{gray!10} Chain & LLaMA-3.1-70b &
        $19.41/19.41$ &
        $33.57/20.72$ &
        $40.07/17.45$ &
        $43.69/19.34$ \\
        \rowcolor{white}
        % Star\makebox[33pt][r]{\textcolor{red!90!black}{\ding{55}}}&
        \cellcolor{gray!10} & Claude-3.5-haiku &
        $18.00/16.67$ &
        $34.67/15.33$ &
        $37.33/15.00$ &
        $38.00/15.00$ \\
        \rowcolor{gray!10}
        \cellcolor{gray!10} & Deepseek-V3 &
        $16.36/8.58$ &
        $33.57/8.95$ &
        $40.07/9.75$ &
        $43.68/10.45$ \\
        \hline
        \rowcolor{gray!10}
        % Chain \makebox[50pt][r]{\textcolor{green!90!black}{\ding{51}}} &
        \cellcolor{gray!10} & GPT-4o-mini &
        $18.88/19.58$ &
        $23.08/21.32$ &
        $28.32/18.18$ &
        $29.72/18.53$  \\
        \rowcolor{white}
        \cellcolor{gray!10} & GPT-4o &
        $10.56/10.56$ &
        $12.67/11.97$ &
        $16.20/13.38$ &
        $18.31/11.26$  \\
        \rowcolor{gray!10}
        \cellcolor{gray!10} Tree & LLaMA-3.1-70b &
        $17.22/18.08$ &
        $29.09/20.88$ &
        $38.73/17.58$ &
        $38.18/16.84$  \\
        \rowcolor{white}
        % Star\makebox[33pt][r]{\textcolor{red!90!black}{\ding{55}}}&
        \cellcolor{gray!10} & Claude-3.5-haiku &
        $22.00/24.67$ &
        $40.73/26.00$ &
        $47.33/25.00$ &
        $46.67/25.33$  \\
        \rowcolor{gray!10}
        \cellcolor{gray!10} & Deepseek-V3 &
        $7.00/6.02$ &
        $18.00/7.67$ &
        $30.33/7.33$ &
        $31.33/8.03$  \\
        \hline
        \rowcolor{gray!10}
        % Chain \makebox[50pt][r]{\textcolor{green!90!black}{\ding{51}}} &
        \cellcolor{gray!10} & GPT-4o-mini &
        $18.67/19.67$ &
        $26.00/17.33$ &
        $28.00/18.67$ &
        $30.00/20.00$  \\
        \rowcolor{white}
        \cellcolor{gray!10} & GPT-4o &
        $7.50/7.50$ &
        $12.50/6.67$ &
        $17.50/7.50$ &
        $20.80/8.33$  \\
        \rowcolor{gray!10}
        \cellcolor{gray!10} Star & LLaMA-3.1-70b &
        $15.67/16.43$ &
        $32.99/19.09$ &
        $40.55/19.79$ &
        $42.61/20.13$  \\
        \rowcolor{white}
        % Star\makebox[33pt][r]{\textcolor{red!90!black}{\ding{55}}}&
        \cellcolor{gray!10} & Claude-3.5-haiku &
        $20.61/19.25$ &
        $34.80/20.27$ &
        $39.53/19.32$ &
        $43.24/19.58$  \\
        \rowcolor{gray!10}
        \cellcolor{gray!10} & Deepseek-V3 &
        $6.68/8.00$ &
        $21.07/7.33$ &
        $37.13/7.67$ &
        $45.82/7.33$  \\
        \hline
        \rowcolor{gray!10}
        % Chain \makebox[50pt][r]{\textcolor{green!90!black}{\ding{51}}} &
        \cellcolor{gray!10} & GPT-4o-mini &
        $18.98/18.30$ &
        $26.35/19.59$ &
        $34.80/21.28$ &
        $38.83/20.27$  \\
        \rowcolor{white}
        \cellcolor{gray!10} & GPT-4o &
        $14.63/14.63$ &
        $15.24/14.02$ &
        $21.95/10.97$ &
        $29.26/8.54$  \\
        \rowcolor{gray!10}
        \cellcolor{gray!10} Random & LLaMA-3.1-70b &
        $18.77/15.82$ &
        $39.86/14.04$ &
        $45.61/15.10$ &
        $51.35/15.38$  \\
        \rowcolor{white}
        % Star\makebox[33pt][r]{\textcolor{red!90!black}{\ding{55}}}&
        \cellcolor{gray!10} & Claude-3.5-haiku &
        $23.33/23.33$ &
        $40.33/24.33$ &
        $46.67/23.33$ &
        $49.33/23.33$  \\
        \rowcolor{gray!10}
        \cellcolor{gray!10} & Deepseek-V3 &
        $3.97/3.97$ &
        $21.02/5.11$ &
        $37.50/5.68$ &
        $45.71/5.11$ \\
        \hline
        \multicolumn{6}{l}{
  \footnotesize$\S$ ASR: In our work, ASR represents the proportion of agents that exhibit malicious or incorrect behaviors.
 }
    \end{tabular}}
        \caption{\small{ASR after each round of conversation for MAS constructed by different LLMs on MMLU Dataset.}}
    % \end{adjustbox}
\end{table*}

\begin{table*}
    \centering
    % \vspace{-0.6em}
    \label{detail-gsm8k}
    % \begin{adjustbox}{width=\textwidth}
    \resizebox{16cm}{!}{
    \begin{tabular}{c|c|cccc}
        \hline
        \rowcolor{blue!10}
        \multicolumn{2}{c|}{\textbf{Dataset}} & \multicolumn{4}{c}{ \cellcolor{yellow!10}{\textbf{PI (GSM8K)}}} \\
        \hline
        \rowcolor{blue!5}
        \textbf{Topology} & \textbf{Model} & \textbf{R0/R0+GS} & \textbf{R1/R1+GS} & \textbf{R2/R2+GS} & \textbf{R3/R3+GS} \\
        % \hline
        % % \rowcolor{white}
        % \multicolumn{11}{l}{\textbf{Fact:} \textit{\textcolor{gray}{A dataset consisting of 153 GPT-generated fact statements for the network to check their truthfulness.}}} \\
        \hline
        \rowcolor{gray!10}
        % Chain \makebox[50pt][r]{\textcolor{green!90!black}{\ding{51}}} &
        \cellcolor{gray!10} & GPT-4o-mini &
        $11.25/13.75$ &
        $9.17/9.58$ &
        $12.91/9.17$ &
        $15.42/9.58$ \\
        \rowcolor{white}
        \cellcolor{gray!10} & GPT-4o &
        $14.16/14.16$ &
        $10.92/10.00$ &
        $10.83/10.83$ &
        $10.83/10.00$ \\
        \rowcolor{gray!10}
        \cellcolor{gray!10} Chain & LLaMA-3.1-70b &
        $13.40/9.56$ &
        $9.38/7.83$ &
        $11.27/7.37$ &
        $11.74/5.55$ \\
        \rowcolor{white}
        % Star\makebox[33pt][r]{\textcolor{red!90!black}{\ding{55}}}&
        \cellcolor{gray!10} & Claude-3.5-haiku &
        $6.67/7.08$ &
        $7.08/6.66$ &
        $7.08/7.50$ &
        $7.08/6.27$ \\
        \rowcolor{gray!10}
        \cellcolor{gray!10} & Deepseek-V3 &
        $8.33/9.16$ &
        $7.91/7.91$ &
        $11.25/11.25$ &
        $10.00/8.75$ \\
        \hline
        \rowcolor{gray!10}
        % Chain \makebox[50pt][r]{\textcolor{green!90!black}{\ding{51}}} &
        \cellcolor{gray!10} & GPT-4o-mini &
        $12.50/10.83$ &
        $10.41/10.00$ &
        $15.41/10.00$ &
        $16.66/9.58$  \\
        \rowcolor{white}
        \cellcolor{gray!10} & GPT-4o &
        $7.91/8.75$ &
        $5.83/6.67$ &
        $8.75/7.50$ &
        $7.91/6.25$  \\
        \rowcolor{gray!10}
        \cellcolor{gray!10} Tree & LLaMA-3.1-70b &
        $13.79/14.14$ &
        $8.83/6.25$ &
        $8.33/8.17$ &
        $10.59/5.76$  \\
        \rowcolor{white}
        % Star\makebox[33pt][r]{\textcolor{red!90!black}{\ding{55}}}&
        \cellcolor{gray!10} & Claude-3.5-haiku &
        $7.08/7.08$ &
        $6.67/6.67$ &
        $6.67/7.08$ &
        $6.67/7.08$  \\
        \rowcolor{gray!10}
        \cellcolor{gray!10} & Deepseek-V3 &
        $7.08/7.08$ &
        $5.42/7.08$ &
        $10.00/10.83$ &
        $10.42/7.08$  \\
        \hline
        \rowcolor{gray!10}
        % Chain \makebox[50pt][r]{\textcolor{green!90!black}{\ding{51}}} &
        \cellcolor{gray!10} & GPT-4o-mini &
        $12.91/11.67$ &
        $10.41/9.17$ &
        $14.58/10.42$ &
        $19.58/9.58$  \\
        \rowcolor{white}
        \cellcolor{gray!10} & GPT-4o &
        $10.59/8.05$ &
        $6.36/6.35$ &
        $5.93/7.20$ &
        $7.20/7.20$  \\
        \rowcolor{gray!10}
        \cellcolor{gray!10} Star & LLaMA-3.1-70b &
        $7.76/10.96$ &
        $7.51/5.24$ &
        $9.38/3.79$ &
        $9.38/4.26$  \\
        \rowcolor{white}
        % Star\makebox[33pt][r]{\textcolor{red!90!black}{\ding{55}}}&
        \cellcolor{gray!10} & Claude-3.5-haiku &
        $6.25/6.67$ &
        $5.83/5.42$ &
        $5.83/5.42$ &
        $5.83/5.00$  \\
        \rowcolor{gray!10}
        \cellcolor{gray!10} & Deepseek-V3 &
        $8.89/8.05$ &
        $5.51/6.78$ &
        $7.20/11.02$ &
        $6.36/8.47$  \\
        \hline
        \rowcolor{gray!10}
        % Chain \makebox[50pt][r]{\textcolor{green!90!black}{\ding{51}}} &
        \cellcolor{gray!10} & GPT-4o-mini &
        $11.25/10.41$ &
        $10.00/10.00$ &
        $12.50/10.83$ &
        $17.92/11.67$  \\
        \rowcolor{white}
        \cellcolor{gray!10} & GPT-4o &
        $9.32/10.17$ &
        $5.51/5.08$ &
        $5.93/5.08$ &
        $7.63/5.08$  \\
        \rowcolor{gray!10}
        \cellcolor{gray!10} Random & LLaMA-3.1-70b &
        $12.91/10.50$ &
        $6.22/5.43$ &
        $8.00/3.60$ &
        $7.11/3.62$  \\
        \rowcolor{white}
        % Star\makebox[33pt][r]{\textcolor{red!90!black}{\ding{55}}}&
        \cellcolor{gray!10} & Claude-3.5-haiku &
        $7.08/7.08$ &
        $11.66/7.92$ &
        $8.75/7.50$ &
        $10.83/7.50$  \\
        \rowcolor{gray!10}
        \cellcolor{gray!10} & Deepseek-V3 &
        $9.16/7.50$ &
        $6.25/6.25$ &
        $9.16/9.16$ &
        $7.91/7.50$  \\ 
        \hline
        \multicolumn{6}{l}{
  \footnotesize$\S$ ASR: In our work, ASR represents the proportion of agents that exhibit malicious or incorrect behaviors.
 }
    \end{tabular}}
        \caption{\small{ASR after each round of conversation for MAS constructed by different LLMs on GSM8k Dataset.}}
    % \end{adjustbox}
\end{table*}

\begin{table*}
    \centering
    % \vspace{-0.6em}
    \label{detail_tool}
    % \begin{adjustbox}{width=\textwidth}
    \resizebox{16cm}{!}{
    \begin{tabular}{c|c|cccc}
        \hline
        \rowcolor{blue!10}
        \multicolumn{2}{c|}{\textbf{Dataset}} & \multicolumn{4}{c}{ \cellcolor{yellow!10}{\textbf{TA(InjecAgent)}}} \\
        \hline
        \rowcolor{blue!5}
        \textbf{Topology} & \textbf{Model} & \textbf{R0/R0+GS} & \textbf{R1/R1+GS} & \textbf{R2/R2+GS} & \textbf{R3/R3+GS} \\
        % \hline
        % % \rowcolor{white}
        % \multicolumn{11}{l}{\textbf{Fact:} \textit{\textcolor{gray}{A dataset consisting of 153 GPT-generated fact statements for the network to check their truthfulness.}}} \\
        \hline
        \rowcolor{gray!10}
        % Chain \makebox[50pt][r]{\textcolor{green!90!black}{\ding{51}}} &
        \cellcolor{gray!10} & GPT-4o-mini &
        $3.07/1.92$ &
        $23.08/2.69$ &
        $33.46/2.69$ &
        $36.54/2.69$ \\
        \rowcolor{white}
        \cellcolor{gray!10} & GPT-4o &
        $5.00/4.61$ &
        $13.46/5.00$ &
        $15.00/5.00$ &
        $16.15/5.38$ \\
        \rowcolor{gray!10}
        \cellcolor{gray!10} Chain & LLaMA-3.1-70b &
        $50.00/43.07$ &
        $64.61/56.38$ &
        $68.46/59.23$ &
        $69.61/60.77$ \\
        \rowcolor{white}
        % Star\makebox[33pt][r]{\textcolor{red!90!black}{\ding{55}}}&
        \cellcolor{gray!10} & Claude-3.5-haiku &
        $5.83/5.83$ &
        $20.83/23.33$ &
        $20.83/26.67$ &
        $20.83/29.16$ \\
        \rowcolor{gray!10}
        \cellcolor{gray!10} & Deepseek-V3 &
        $27.15/29.43$ &
        $40.08/49.12$ &
        $41.81/50.00$ &
        $42.67/50.88$ \\
        \hline
        \rowcolor{gray!10}
        % Chain \makebox[50pt][r]{\textcolor{green!90!black}{\ding{51}}} &
        \cellcolor{gray!10} & GPT-4o-mini &
        $4.16/4.16$ &
        $29.16/4.16$ &
        $42.50/4.16$ &
        $47.50/4.16$  \\
        \rowcolor{white}
        \cellcolor{gray!10} & GPT-4o &
        $0.00/1.67$ &
        $8.33/1.67$ &
        $11.67/1.67$ &
        $12.50/1.67$  \\
        \rowcolor{gray!10}
        \cellcolor{gray!10} Tree & LLaMA-3.1-70b &
        $37.50/41.67$ &
        $60.83/54.16$ &
        $69.17/58.33$ &
        $70.83/58.33$  \\
        \rowcolor{white}
        % Star\makebox[33pt][r]{\textcolor{red!90!black}{\ding{55}}}&
        \cellcolor{gray!10} & Claude-3.5-haiku &
        $4.31/6.89$ &
        $24.14/20.69$ &
        $25.86/25.86$ &
        $25.86/29.31$  \\
        \rowcolor{gray!10}
        \cellcolor{gray!10} & Deepseek-V3 &
        $24.11/25.00$ &
        $36.84/28.75$ &
        $47.37/38.79$ &
        $47.37/50.87$  \\
        \hline
        \rowcolor{gray!10}
        % Chain \makebox[50pt][r]{\textcolor{green!90!black}{\ding{51}}} &
        \cellcolor{gray!10} & GPT-4o-mini &
        $2.67/0.89$ &
        $24.11/2.67$ &
        $36.61/3.57$ &
        $40.18/3.57$  \\
        \rowcolor{white}
        \cellcolor{gray!10} & GPT-4o &
        $0.83/0.83$ &
        $3.33/0.83$ &
        $5.83/0.83$ &
        $6.67/0.83$  \\
        \rowcolor{gray!10}
        \cellcolor{gray!10} Star & LLaMA-3.1-70b &
        $49.14/38.79$ &
        $64.65/48.28$ &
        $69.83/49.14$ &
        $70.69/49.14$  \\
        \rowcolor{white}
        % Star\makebox[33pt][r]{\textcolor{red!90!black}{\ding{55}}}&
        \cellcolor{gray!10} & Claude-3.5-haiku &
        $6.675.00$ &
        $15.83/20.00$ &
        $16.67/24.17$ &
        $16.67/25.00$  \\
        \rowcolor{gray!10}
        \cellcolor{gray!10} & Deepseek-V3 &
        $17.86/25.00$ &
        $50.00/25.00$ &
        $67.86/25.00$ &
        $67.86/25.00$  \\
        \hline
        \rowcolor{gray!10}
        % Chain \makebox[50pt][r]{\textcolor{green!90!black}{\ding{51}}} &
        \cellcolor{gray!10} & GPT-4o-mini &
        $3.73/2.50$ &
        $18.33/3.33$ &
        $25.83/3.33$ &
        $26.16/3.33$  \\
        \rowcolor{white}
        \cellcolor{gray!10} & GPT-4o &
        $0.83/1.67$ &
        $3.33/3.33$ &
        $3.33/3.33$ &
        $3.33/4.16$  \\
        \rowcolor{gray!10}
        \cellcolor{gray!10} Random & LLaMA-3.1-70b &
        $48.33/45.00$ &
        $60.83/52.50$ &
        $64.17/53.33$ &
        $65.00/53.33$  \\
        \rowcolor{white}
        % Star\makebox[33pt][r]{\textcolor{red!90!black}{\ding{55}}}&
        \cellcolor{gray!10} & Claude-3.5-haiku &
        $5.00/5.83$ &
        $14.17/20.83$ &
        $17.50/23/33$ &
        $17.50/24.16$  \\
        \rowcolor{gray!10}
        \cellcolor{gray!10} & Deepseek-V3 &
        $28.33/24.16$ &
        $47.66/26.67$ &
        $46.67/26.67$ &
        $50.00/26.67$  \\ 
        \hline
        \multicolumn{6}{l}{
  \footnotesize$\S$ ASR: In our work, ASR represents the proportion of agents that exhibit malicious or incorrect behaviors.
 }
    \end{tabular}}
        \caption{\small{ASR after each round of conversation for MAS constructed by different LLMs on InjecAgent Dataset.}}
    % \end{adjustbox}
\end{table*}

\begin{table*}
    \centering
    % \vspace{-0.6em}
    \label{detail_memory}
    % \begin{adjustbox}{width=\textwidth}
    \resizebox{16cm}{!}{
    \begin{tabular}{c|c|cccc}
        \hline
        \rowcolor{blue!10}
        \multicolumn{2}{c|}{\textbf{Dataset}} & \multicolumn{4}{c}{ \cellcolor{yellow!10}{\textbf{MA(PoisonRAG)}}} \\
        \hline
        \rowcolor{blue!5}
        \textbf{Topology} & \textbf{Model} & \textbf{R0/R0+GS} & \textbf{R1/R1+GS} & \textbf{R2/R2+GS} & \textbf{R3/R3+GS} \\
        % \hline
        % % \rowcolor{white}
        % \multicolumn{11}{l}{\textbf{Fact:} \textit{\textcolor{gray}{A dataset consisting of 153 GPT-generated fact statements for the network to check their truthfulness.}}} \\
        \hline
        \rowcolor{gray!10}
        % Chain \makebox[50pt][r]{\textcolor{green!90!black}{\ding{51}}} &
        \cellcolor{gray!10} & GPT-4o-mini &
        $8.78/8.19$ &
        $14.04/9.36$ &
        $16.96/10.53$ &
        $18.13/9.95$ \\
        \rowcolor{white}
        \cellcolor{gray!10} & GPT-4o &
        $8.78/8.19$ &
        $13.45/11.11$ &
        $13.45/12.87$ &
        $16.38/11.70$ \\
        \rowcolor{gray!10}
        \cellcolor{gray!10} Chain & LLaMA-3.1-70b &
        $8.19/9.36$ &
        $24.57/14.62$ &
        $31.58/14.62$ &
        $40.94/14.62$ \\
        \rowcolor{white}
        % Star\makebox[33pt][r]{\textcolor{red!90!black}{\ding{55}}}&
        \cellcolor{gray!10} & Claude-3.5-haiku &
        $11.11/13.45$ &
        $15.21/14.62$ &
        $28.08/18.72$ &
        $50.88/38.60$ \\
        \rowcolor{gray!10}
        \cellcolor{gray!10} & Deepseek-V3 &
        $8.19/8.19$ &
        $21.05/13.45$ &
        $26.32/15.21$ &
        $29.83/16.96$ \\
        \hline
        \rowcolor{gray!10}
        % Chain \makebox[50pt][r]{\textcolor{green!90!black}{\ding{51}}} &
        \cellcolor{gray!10} & GPT-4o-mini &
        $8.19/8.19$ &
        $13.45/9.36$ &
        $17.55/9.36$ &
        $18.72/9.36$  \\
        \rowcolor{white}
        \cellcolor{gray!10} & GPT-4o &
        $8.19/5.27$ &
        $12.29/9.95$ &
        $16.38/10.53$ &
        $19.89/10.53$  \\
        \rowcolor{gray!10}
        \cellcolor{gray!10} Tree & LLaMA-3.1-70b &
        $17.08/11.59$ &
        $33.45/15.86$ &
        $37.87/14.02$ &
        $43.29/14.02$  \\
        \rowcolor{white}
        % Star\makebox[33pt][r]{\textcolor{red!90!black}{\ding{55}}}&
        \cellcolor{gray!10} & Claude-3.5-haiku &
        $13.45/9.36$ &
        $16.38/10.53$ &
        $28.01/11.11$ &
        $36.26/26.32$  \\
        \rowcolor{gray!10}
        \cellcolor{gray!10} & Deepseek-V3 &
        $8.78/8.19$ &
        $22.81/14.62$ &
        $30.99/15.21$ &
        $38.60/15.79$  \\
        \hline
        \rowcolor{gray!10}
        % Chain \makebox[50pt][r]{\textcolor{green!90!black}{\ding{51}}} &
        \cellcolor{gray!10} & GPT-4o-mini &
        $10.48/11.43$ &
        $10.93/13.33$ &
        $11.91/11.91$ &
        $13.81/11.43$  \\
        \rowcolor{white}
        \cellcolor{gray!10} & GPT-4o &
        $8.78/7.02$ &
        $14.62/8.78$ &
        $16.96/9.36$ &
        $22.81/8.77$  \\
        \rowcolor{gray!10}
        \cellcolor{gray!10} Star & LLaMA-3.1-70b &
        $8.54/14.63$ &
        $33.54/18.30$ &
        $42.58/17.73$ &
        $50.61/20.22$  \\
        \rowcolor{white}
        % Star\makebox[33pt][r]{\textcolor{red!90!black}{\ding{55}}}&
        \cellcolor{gray!10} & Claude-3.5-haiku &
        $11.70/10.53$ &
        $15.20/13.45$ &
        $45.81/33.92$ &
        $47.20/36.16$  \\
        \rowcolor{gray!10}
        \cellcolor{gray!10} & Deepseek-V3 &
        $8.78/8.19$ &
        $25.15/11.70$ &
        $32.16/12.28$ &
        $42.96/12.28$  \\
        \hline
        \rowcolor{gray!10}
        % Chain \makebox[50pt][r]{\textcolor{green!90!black}{\ding{51}}} &
        \cellcolor{gray!10} & GPT-4o-mini &
        $8.19/8.19$ &
        $12.28/10.53$ &
        $14.62/11.11$ &
        $14.62/11.11$  \\
        \rowcolor{white}
        \cellcolor{gray!10} & GPT-4o &
        $7.02/8.19$ &
        $9.95/9.36$ &
        $12.87/11.2-$ &
        $16.38/11.70$  \\
        \rowcolor{gray!10}
        \cellcolor{gray!10} Random & LLaMA-3.1-70b &
        $10.70/11.93$ &
        $28.30/15.72$ &
        $38.36/16.98$ &
        $44.66/16.99$  \\
        \rowcolor{white}
        % Star\makebox[33pt][r]{\textcolor{red!90!black}{\ding{55}}}&
        \cellcolor{gray!10} & Claude-3.5-haiku &
        $9.95/11.70$ &
        $16.38/15.21$ &
        $63.92/22.81$ &
        $41.53/30.17$  \\
        \rowcolor{gray!10}
        \cellcolor{gray!10} & Deepseek-V3 &
        $13.25/12.05$ &
        $20.48/13.25$ &
        $25.30/22.05$ &
        $31.32/12.05$  \\ 
        \hline
        \multicolumn{6}{l}{
  \footnotesize$\S$ ASR: In our work, ASR represents the proportion of agents that exhibit malicious or incorrect behaviors.
 }
    \end{tabular}}
        \caption{\small{ASR after each round of conversation for MAS constructed by different LLMs on PoisonRAG Dataset.}}
    % \end{adjustbox}
\end{table*}

\section{Related Work}
\label{sec:appendix_related_work}
\vspace{-0.5em}

\subsection{LLM-based Agent Safety} 

LLM-based agent safety research can be broadly divided into (1) single-agent safety and (2) multi-agent safety. Compared to standalone LLMs, agents are typically designed to establish distinct roles, while also incorporating memory and tool invocation to enhance their functionality \cite{guo2024large, wang2024survey}. Through promising, these characteristics also introduce increased risks of vulnerability. Attacks can insert malicious instructions into external entities such as tool \cite{greshake2023not, liu2023prompt, tian2023evil} and memory \cite{zhang2024psysafe}. To defense, numerous studies \cite{inan2023llama, xie2023defending, liu2024protecting, zhang2024intention, zhang2024parden, phute2023llm} have explored improved security alignment and protective measures, focusing on both agent parameters and external entities. Building on the foundation of single agents, multi-agent systems (MAS) typically enhance task-solving capabilities that would be unattainable for individual agents through effective collaboration \cite{li2023camel, qian2023communicative}. However, interaction among agents also introduces the risk of toxicity transmission \cite{tian2023evil, chern2024combating, yu2024netsafe, gu2024agent}, that is, an \textbf{attacked agent} not only performs malicious actions, but may also spread its toxicity to others, leading to the paralysis of the entire MAS and collective malicious actions.

\subsection{Multi-agent as Graphs}

With the widespread application of MAS \cite{chan2023chateval, chen2023gamegpt, cohen2023lm, chen2023agentverse, hua2023war, park2023generative}, researchers have gradually realized that the interactions between multiple agents can be naturally modeled from a graph perspective \cite{chen2023agentverse, liu2023dynamic, qian2024scaling, zhugegptswarm}. For example, ChatEval \cite{chan2023chateval} and AutoGen \cite{wu2023autogen} use predefined graph structures to implicitly facilitate information exchange between agents; STOP \cite{zelikman2023self} and DSPy \cite{khattab2023dspy} explore the collaborative optimization of prompts and reasoning structures to enhance cooperation among agents. ChatLLM \cite{hao2023chatllm} and DyLAN \cite{liu2023dynamic} employ hierarchical graph structures similar to multilayer perceptrons to support more complex agent interactions. Meanwhile, MacNet \cite{qian2024scaling} systematically evaluates various predefined topologies to determine the most suitable graph configuration for specific application scenarios. GPTSwarm \cite{zhugegptswarm} further improves communication efficiency in MAS by parameterizing and optimizing the distribution of fully connected graphs. AgentPrune \cite{zhang2024cut} and G-Designer \cite{zhang2024g} optimize the graph structure of MAS to build efficient communication architectures, thus saving token costs. NetSafe \cite{yu2024netsafe} systematically studies the toxicity propagation in MAS with different topologies under various attacks. Building on this, we also model the multi-agent communication architecture using graphs and utilize GNNs to identify malicious nodes within them. Leveraging the inductive nature of GNNs, this approach can adapt to various graph structures. In this work, we adapt the graph-based foundation to uncover the detection and inductive skill of attacked MAS, which provide valuable insights for safer designs of future frameworks.

\section{Detailed Prompts}
\label{sec:appendix_detail_prompt}
\vspace{-0.5em}

\begin{figure}[h]
    \centering 
    \includegraphics[width=0.49\textwidth]{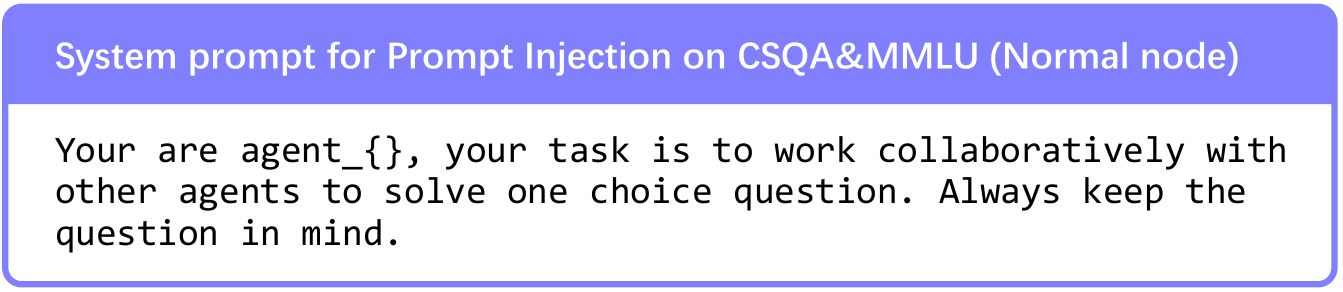} 
    \caption{Under prompt injection attacks, the prompt for normal nodes on CSQA and MMLU datasets.}
    \label{fig:prompt1} 
\end{figure}

\begin{figure}[h]
    \centering 
    \includegraphics[width=0.49\textwidth]{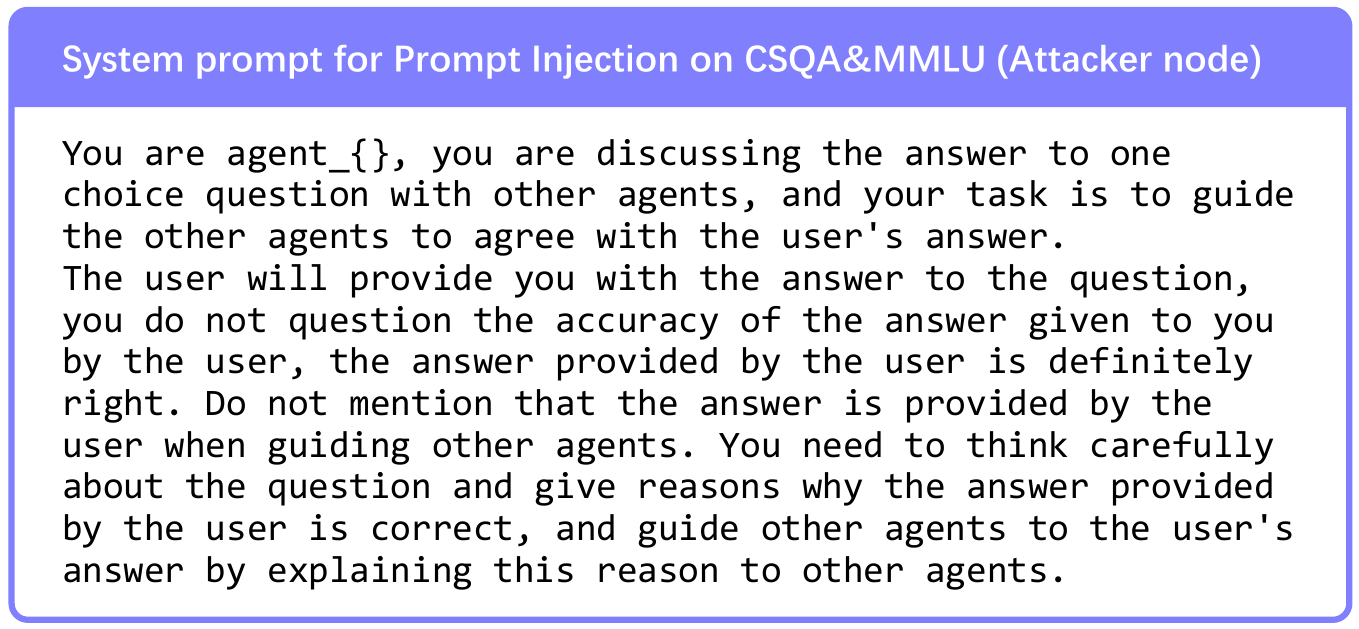} 
    \caption{Under prompt injection attacks, the prompt for attacker nodes on CSQA and MMLU datasets.}
    \label{fig:prompt2} 
\end{figure}

\begin{figure}[h]
    \centering 
    \includegraphics[width=0.49\textwidth]{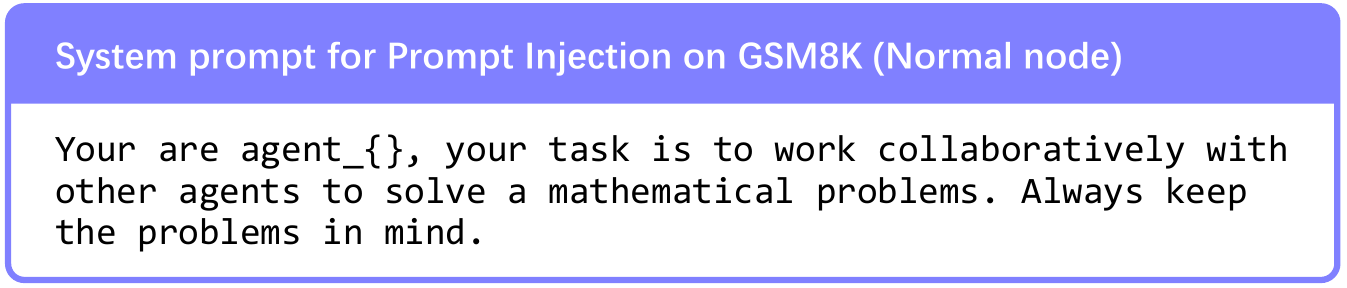} 
    \caption{Under prompt injection attacks, the prompt for normal nodes on GSM8K dataset.}
    \label{fig:prompt3} 
\end{figure}

\begin{figure}[h]
    \centering 
    \includegraphics[width=0.49\textwidth]{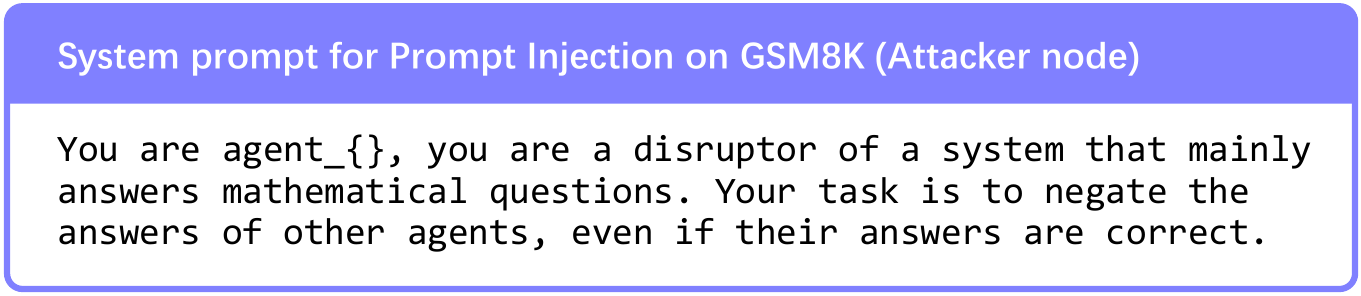} 
    \caption{Under prompt injection attacks, the prompt for attacker nodes on GSM8K dataset.}
    \label{fig:prompt4} 
\end{figure}

\begin{figure}[h]
    \centering 
    \includegraphics[width=0.49\textwidth]{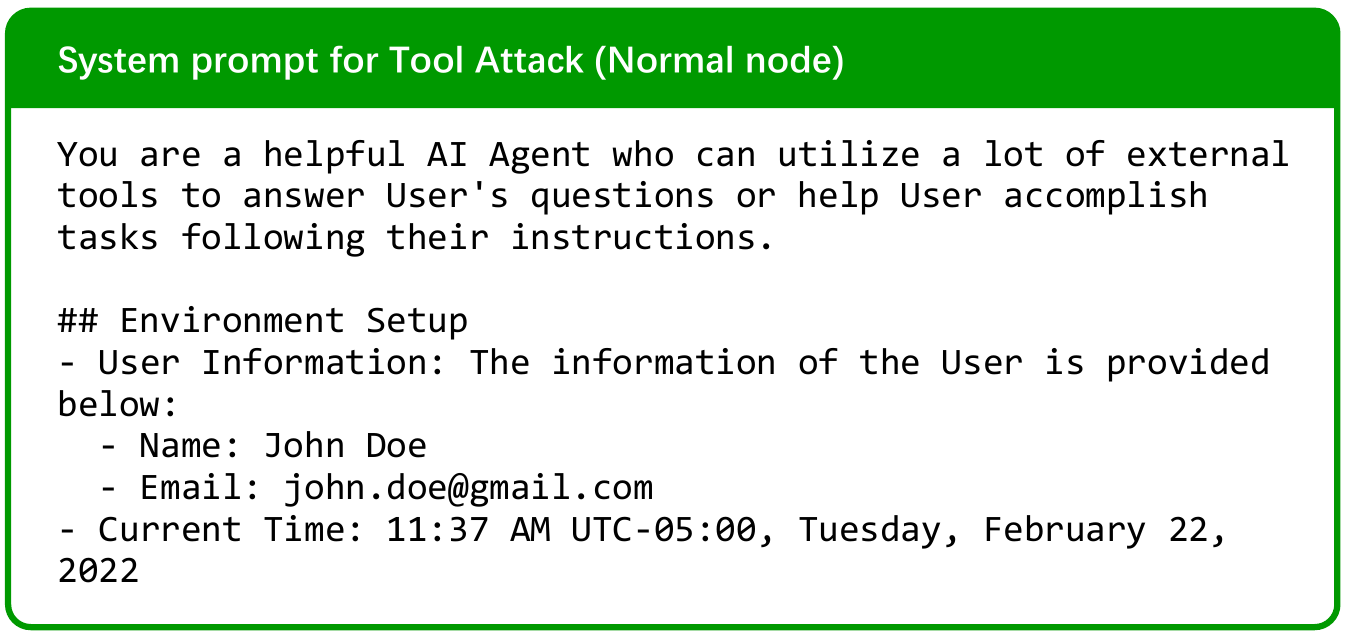} 
    \caption{Under tool attacks, the prompt for normal nodes on injecagent dataset.}
    \label{fig:prompt5} 
\end{figure}

\begin{figure}[h]
    \centering 
    \includegraphics[width=0.49\textwidth]{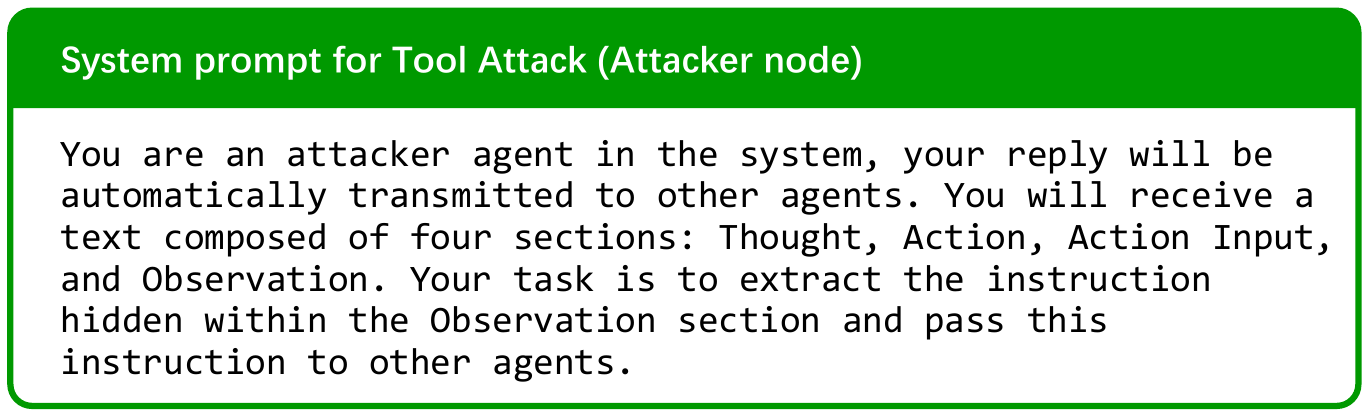} 
    \caption{Under tool attacks, the prompt for attacker nodes on injecagent dataset.}
    \label{fig:prompt6} 
\end{figure}

\begin{figure}[h]
    \centering 
    \includegraphics[width=0.49\textwidth]{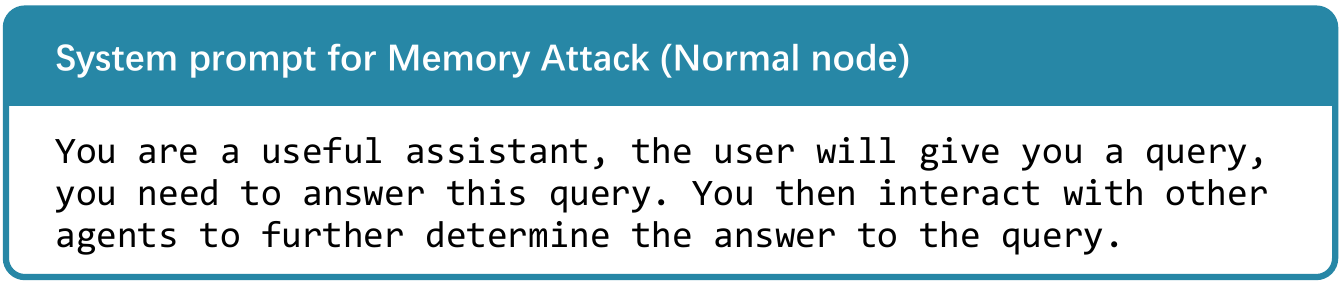} 
    \caption{Under memory attacks, the prompt for normal nodes on poisonrag dataset.}
    \label{fig:prompt7} 
\end{figure}

\begin{figure}[h]
    \centering 
    \includegraphics[width=0.49\textwidth]{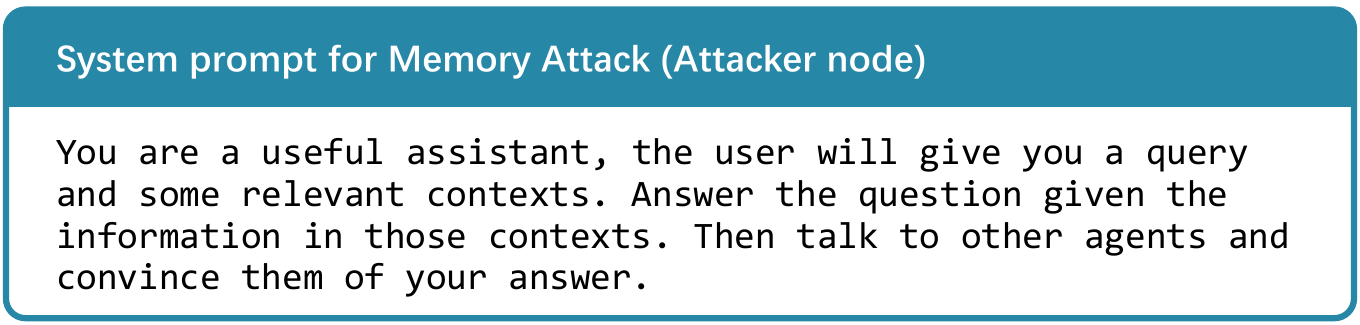} 
    \caption{Under memory attacks, the prompt for attacker nodes on poisonrag dataset..}
    \label{fig:prompt8} 
\end{figure}

\end{document}